\documentclass[12pt]{article}

\usepackage{scicite}
\usepackage{times}

\newenvironment{sciabstract}{%
\begin{quote} \bf}
{\end{quote}}

\newcounter{lastnote}

\usepackage{xparse}
\usepackage{xspace}
\usepackage{color}
\usepackage{todonotes}

\usepackage{booktabs}
\usepackage{multirow}

\usepackage{textcomp}

\usepackage{enumerate}

\usepackage{authblk}

\usepackage{braket}

\usepackage{amsmath, amssymb}
\usepackage{mathtools}

\DeclareDocumentCommand{\fm}{ m }{\ifmmode #1 \else $#1$\fi}

\def\benine{\fm{^{9}\mathrm{Be}^{+}}\xspace}

\def\mgfive{\fm{^{25}\mathrm{Mg}^{+}}\xspace}

\def\beshort{\fm{\mathrm{B}}\xspace}
\def\mgshort{\fm{\mathrm{M}}\xspace}

\def\besubs{\fm{\mathrm{B}}\xspace}
\def\mgsubs{\fm{\mathrm{M}}\xspace}

\def\bebrstate{\fm{\ket{\mathrm{Bright}}_\besubs}\xspace}
\def\bedkstate{\fm{\ket{\mathrm{Dark}}_\besubs}\xspace}

\def\mgbrstate{\fm{\ket{\mathrm{Bright}}_\mgsubs}\xspace}
\def\mgdkstate{\fm{\ket{\mathrm{Dark}}_\mgsubs}\xspace}

\def\bmmbnotation{\beone--\mgone--\mgtwo--\betwo}

\def\fouriontxt{four-ion chain\xspace}

\def\bmpair{\beshort-and-\mgshort}

\def\beone{\fm{\mathrm{B}_\mathrm{1}}\xspace}
\def\betwo{\fm{\mathrm{B}_\mathrm{2}}\xspace}
\def\mgone{\fm{\mathrm{M}_\mathrm{1}}\xspace}
\def\mgtwo{\fm{\mathrm{M}_\mathrm{2}}\xspace}

\def\ustate{\fm{\ket{\uparrow}}\xspace}
\def\dstate{\fm{\ket{\downarrow}}\xspace}

\def\p{\fm{+}\xspace}
\def\r{\fm{r}\xspace}

\def\bellstaters{\fm{\frac{1}{\sqrt{2}}(\ket{\uparrow\uparrow}_\mgsubs + \ket{\downarrow\downarrow}_\mgsubs)}\xspace}

\def\phiplusmg{\fm{\ket{\Phi^\mathrm{+}}_\mgsubs}\xspace}

\newcommand{\cE}{\mathcal{E}}
\newcommand{\cL}{\mathcal{L}}
\newcommand{\cS}{\mathcal{S}}
\newcommand{\Tr}{\mathrm{Tr}}

\def\cnot{\fm{\mathrm{CNOT}}\xspace}
\def\cnottxt{CNOT\xspace}
\def\msfull{M{\o}lmer-S{\o}rensen\xspace}
\def\msgate{MS\xspace}

\def\singlewell{{single well}\xspace}

\def\nbar{\fm{\bar{n}}\xspace}

\def\us{\fm{\mu\mathrm{s}}\xspace}
\def\um{\fm{\mu\mathrm{m}}\xspace}

\DeclareDocumentCommand{\tfinmhz}{ m }{\fm{#1\,\mathrm{MHz}}\xspace}

\def\paratrapandqubit{Trap and Qubit\xspace}
\def\paracp{Crystal Preparation\xspace}
\def\parashuttling{Shuttling Operations\xspace}
\def\paraspam{State Preparation and Detection\xspace}
\def\parasbc{Sideband Cooling\xspace}
\def\paragates{Logic Gates\xspace}
\def\parapt{Phase Tracking\xspace}
\def\paradc{Drifts and Calibration\xspace}
\def\paratomography{Quantum Process Tomography\xspace}
\def\paraptm{Pauli Transfer Matrix\xspace}
\def\paralrt{Likelihood Ratio Test\xspace}
\def\paraes{Depolarizing Error Model\xspace}

\title{Quantum gate teleportation between separated qubits in a trapped-ion processor}

\author{
Yong Wan,$^{1,2\ast\dagger}$
Daniel Kienzler,$^{1,2\dagger\ddagger}$
Stephen D. Erickson,$^{1,2}$
Karl H. Mayer,$^{1, 2}$
Ting Rei Tan,$^{1, 2\ddagger}$
Jenny J. Wu,$^{1,2}$
Hilma M. Vasconcelos,$^{1,2,4}$
Scott Glancy,$^{1}$
Emanuel Knill,$^{1}$
David J. Wineland,$^{1,2,3}$
Andrew C. Wilson,$^{1}$
Dietrich Leibfried$^{1}$}

\affil{$^{1}$National Institute of Standards and Technology, 325 Broadway, Boulder, CO 80305,
USA\\
$^{2}$Department of Physics, University of Colorado, Boulder, CO 80309, USA\\
$^{3}$Department of Physics, University of Oregon, Eugene, OR 97403, USA\\
$^{4}$Departamento de Engenharia de Teleinform\'atica, Universidade Federal do Cear\'a, Fortaleza, Brazil\\
$^\ast$To whom correspondence should be addressed. E-mail: yong.wan@nist.gov. \\
$^{\dagger}$These authors contributed equally to this work.
$^{\ddagger}$Current address: ETH Zurich, Otto-Stern-Weg 1, HPF E10, 8093 Zurich, Switzerland
(D.K.); Department of Physics, National University of Singapore, 2 Science Drive 3, 117551 Singapore (T.R.T); Centre for Quantum Technologies, 3 Science Drive 2, 117543 Singapore (T.R.T.).\\
}

\date{}

\begin{document}

\baselineskip24pt
\maketitle

\def\fidintdstwo{$[0.845, 0.872]$\xspace}
\def\fidintdstwobf{$\mathbf{[0.845, 0.872]}$\xspace}
\def\fidledstwo{$[0.845, 0.888]$\xspace}

\begin{sciabstract}
Large-scale quantum computers will require quantum gate operations between
widely separated qubits.
A method for implementing such operations, known as quantum gate teleportation (QGT),
requires only local operations,
 classical communication, and  shared entanglement.
We demonstrate QGT in a scalable architecture by deterministically teleporting a controlled-NOT (CNOT) gate between two
qubits in spatially separated locations in an ion trap.
The entanglement fidelity of our teleported \cnottxt is in the interval  \fidintdstwobf at the 95\% confidence level.
The implementation combines ion shuttling with individually-addressed single-qubit
rotations and detections, same- and mixed-species two-qubit gates, and real-time conditional operations,
thereby demonstrating essential tools for scaling trapped-ion quantum computers combined in a single device.
\end{sciabstract}

Quantum computers have the potential to solve problems that are intractable using
conventional computers.
However,
many quantum bits (qubits) are required to outperform conventional computing capabilities, and scaling quantum computers to be practically useful is difficult \cite{preskill_quantum_2018}.
As the system size increases, the average distance between
qubits grows, making it harder to connect arbitrary qubits.
Quantum gate teleportation (QGT) is a uniquely quantum solution that enables logical gates between spatially separated qubits, where shared entanglement eliminates the need for a direct quantum coherent interaction \cite{gottesman_demonstrating_1999,eisert_optimal_2000}.

There are several proposals for scaling up to larger numbers of qubits in trapped-ion systems. These include the ``quantum charge-coupled device'' (QCCD) architecture, which
incorporates a large array of segmented electrodes to create  trapping zones specialized for roles such as loading ions, processing, and memory storage
\cite{wineland_experimental_1998,kielpinski_architecture_2002}.
Qubits can interact by being physically moved to the same zone.
A variant of this approach couples different zones by creating entanglement via a photonic network \cite{monroe_large-scale_2014}.
Both approaches will benefit from a way to perform gate operations between separated qubits via QGT, which mitigates latency from transmitting quantum information between  zones, provided  that the required entangled ancilla pairs are prepared and distributed ahead of time during unrelated processor functions. This entanglement can be produced using various methods, including unitary gates, dissipative schemes \cite{lin_dissipative_2013}, and photonic links \cite{moehring_entanglement_2007}.

Progress towards distributed quantum computation has been made with quantum state teleportation \cite{pirandola_advances_2015},
where an arbitrary state is transferred between remote parties \cite{bennett_teleporting_1993, gottesman_demonstrating_1999}.
Using state teleportation, a two-qubit gate between two parties, Alice and Bob,
can be implemented by teleporting Alice's input state to Bob, applying local
two-qubit gates at Bob's location, and teleporting Alice's half of the output back to her.
This process consumes a minimum of two shared entangled pairs.

For a controlled-NOT (\cnot) gate, the task can be achieved more efficiently
using the protocol
depicted in Fig.~1A \cite{eisert_optimal_2000}, without the need to physically bring the qubits together or
teleport the states back and forth. This protocol achieves the minimum possible overhead, requiring only a single entangled pair shared between two locations, local operations, and classical communication.
The protocol implementing a teleported \cnot between qubits \beone and \betwo works as follows. The initial entanglement between qubits \mgone and \mgtwo is transferred to \beone and \mgtwo through the first local \cnot, \mgone detection, and conditional operation on \mgtwo. With the information about \beone's state now shared with \mgtwo, \beone is the effective control of the second local \cnot that acts on \betwo. The remaining operations serve to disentangle \mgtwo from \beone and \betwo, resulting in an effective \cnot between \beone and \betwo.

This type of teleported gate has been demonstrated probabilistically with photonic systems, where
the required conditional operations were implemented with passive optical elements and
post-selection \cite{huang_experimental_2004,gao_teleportation-based_2010}.
More recently, a deterministic \cnot was teleported between two superconducting cavity qubits using an entangled pair of transmons \cite{chou_deterministic_2018}.

Here, we demonstrate a {deterministic} teleported \cnot
between two \benine ions
using a shared entangled pair of \mgfive ions.
It combines key elements for scalable quantum computation with trapped ions, including separation and transport of mixed-species ion crystals, local
same- and mixed-species two-qubit  gates  \cite{tan_multi-element_2015},
individually-addressed single-qubit rotations and detection, and conditional operations
based on measurement results.
We use quantum process tomography (QPT) to characterize the teleported \cnot. We simplify the demonstration by using only one laser interaction zone (LIZ) and transporting the separated qubits to this location, but the key elements of the protocol are retained.

Our experiment uses two \benine ions (\beone, \betwo) and two \mgfive ions (\mgone, \mgtwo) trapped in a segmented linear Paul trap.
The qubits are encoded in the   $\ket{F=1, m_F=1}_\besubs\equiv \ket{\uparrow}_\besubs$ and $\ket{2, 0}_\besubs \equiv \ket{\downarrow}_\besubs$ hyperfine states
of \benine
and the $\ket{2, 0}_\mgsubs \equiv \ket{\uparrow}_\mgsubs$
and $\ket{3, 1}_\mgsubs \equiv \ket{\downarrow}_\mgsubs$ states of \mgfive.
We use the symbol \beshort (\mgshort) for \benine (\mgfive) ions, and label respective states with subscript \beshort (\mgshort).
To begin each experiment, a \fouriontxt is initialized in the order
\beone--\mgone--\mgtwo--\betwo in a potential well where all laser beams pass through the trap, a region we refer to as the LIZ
(Fig.~2).
Qubit-state measurements of \beshort (\mgshort) ions are realized
by state-dependent fluorescence
detection with  $313\,\mathrm{nm}$ ($280\,\mathrm{nm}$) resonant light after
transferring the population from the computational basis  to the measurement basis
$\ustate_\besubs \rightarrow \ket{2, 2}_\besubs \equiv \bebrstate$
and
$\dstate_\besubs \rightarrow \ket{1, -1}_\besubs \equiv \bedkstate$
\big($\ket{\downarrow}_\mgsubs\rightarrow\ket{3, 3}_\mgsubs\equiv\mgbrstate$
and
$\ket{\uparrow}_\mgsubs\rightarrow\ket{2,-2}_\mgsubs\equiv\mgdkstate$\big) \cite{noauthor_materials_nodate}.
Segmented trap electrodes enable the use of time-varying potentials to split the
ion crystal into selected subsets and to transport them
to and from	 the LIZ
\cite{blakestad_high-fidelity_2009,blakestad_near-ground-state_2011}.
Spatial separation enables individual addressing of ions of the same
species, while ions of different species are distinguished by their
well-separated resonant wavelengths.

Stimulated Raman transitions are used for all coherent qubit operations.
A pair of co-propagating laser beams for each species
drives single-qubit rotations $\hat{R}(\theta, \phi)$,
and rotation around the $z$-axis $\hat{R}_\mathrm{Z}(\alpha)$ is implemented by phase-shifting all subsequent single-qubit rotations for that qubit \cite{noauthor_materials_nodate}.
A pair of perpendicular laser beams for each species drives
 two-qubit \msfull (\msgate) entangling gates
\cite{sorensen_quantum_1999}.
Both pairs of Raman beams are applied simultaneously to drive mixed-species \msgate gates
\cite{tan_multi-element_2015}.
We construct $\mathrm{CNOT}_{\mathrm{C}\rightarrow\mathrm{T}}$
($\mathrm{C}$ for control and $\mathrm{T}$ for target)
and the Bell-state-generating gate $\hat{F}$ using single-qubit rotations and \msgate gates
\cite{lee_phase_2005,tan_multi-element_2015,noauthor_materials_nodate}.

The circuit diagram for our  teleported \cnot is shown in Fig.~1B, and
ion configurations during QGT are
illustrated in Fig.~2. After ground-state cooling the \fouriontxt, the algorithm begins with the \beshort and \mgshort ions in \bebrstate and $\ket{\downarrow}_\mgsubs$, respectively. $\hat{F}$ is applied to the two
\mgshort ions to generate the Bell state
$\phiplusmg=\bellstaters$
(Fig. 2A).
Afterwards, the chain is split into two \bmpair pairs in separated regions of a double-well potential
(Fig.~2B) which is translated
 to bring
\beone--\mgone into the LIZ. There we ground-state cool \beone--\mgone by addressing \beone,
prepare \beone to its input state, and apply $\cnot_{\beone\rightarrow\mgone}$.
Then, \mgone is detected (Fig. 2C).
Its qubit state is determined by comparing the number of  detected photons to a preset threshold.
The double-well potential is then translated to move \mgtwo--\betwo into the LIZ, where the pair is ground-state cooled by addressing \betwo.
Qubit \betwo is then prepared in its input state, and the conditional operation $\hat{R}(\pi, 0)$ is applied
on \mgtwo  if \mgone was measured to be in $ \ket{\downarrow}_\mgsubs$.
Next we apply $\cnot_{\mgtwo\rightarrow\betwo}$,
followed by a rotation $\hat{R}(\frac{\pi}{2},-\frac{\pi}{2})$ and detection of \mgtwo.
A rotation selecting the measurement axis for state tomography is applied to \betwo, which is
then mapped out to the measurement basis, but not yet detected (Fig.
2D). This mapping reduces the depumping of \betwo
from stray scattered light when detecting \beone later in the process.
The double-well potential is translated back to bring \beone--\mgone into the LIZ, where we
apply the conditional operation $\hat{R}_\mathrm{Z}(\pi)$ if \mgtwo was
measured to be in  $\ket{\downarrow}_\mgsubs$,
followed by a single-qubit rotation selecting the measurement axis and a
measurement of \beone (Fig.~2E).
Subsequently,  \mgtwo--\betwo are shuttled back into the LIZ where \betwo is detected (Fig.~2F).
At the end of this sequence, the four ions are recombined into a \singlewell to prepare for the next repetition of the experiment.

We used QPT \cite{chuang_prescription_1997} to characterize our teleported \cnot between the two \beshort ions.
$144$ different combinations of input states and measurement axes were implemented in random
order and each for approximately 300 consecutive experiment executions.
Two complete sets of tomography data were acquired.
We developed a protocol for data analysis on Data Set 1 while remaining blind to Data Set 2, and then applied this protocol to Data Set 2.
We summarize the analysis methods and results for Data Set 2 below.

From the observed measurement outcomes,
we determined the most likely quantum process by maximum likelihood (ML) estimation, and inferred a 95\% confidence interval of [0.845, 0.872] for the entanglement fidelity
with
respect
to an ideal \cnot.
The matrix representing the quantum process is shown in Fig.~3.
For details of our analysis, see \cite{noauthor_materials_nodate}.

Ideally, the observed data should be consistent with the assumption of a single quantum process, but  drifts in control parameters on time-scales much slower than a single QGT experiment can lead to imperfections. To detect departure from this assumption,
 we applied a likelihood ratio (LR) test \cite{casella_statistical_nodate,blume-kohout_demonstration_2017}.
An LR was computed from the experimental data and compared to the distribution of LRs obtained from synthetic datasets generated by parametric bootstrapping \cite{efron_introduction_1994}. The test indicated that our data was inconsistent with a single quantum process \cite{noauthor_materials_nodate}.
Motivated by this finding, we discovered drifts in single-qubit-rotation angles that eluded our feedback mechanisms but can be addressed in future experiments \cite{noauthor_materials_nodate}.
We verified through numerical simulation that realistic fluctuations of single-qubit-rotation angles
are capable of causing an inconsistency comparable to that observed in our data.
Although such drifts are not  major sources of infidelity for this experiment \cite{noauthor_materials_nodate}, such consistency checks could be an important diagnostic that can supplement other benchmarking techniques and uncover overlooked sources of infidelity.

We list the dominant error sources and estimate their combined impact in Table~\ref{table:error_budget}.
If all errors are mutually independent, the total error is
$0.16(2)$.
A more accurate description of the impact of individual errors using
a depolarizing model predicts a
process fidelity of $0.88(1)$,
which is near the upper limit of the 95\% ML confidence interval, indicating that the major error
sources are included in the error propagation model \cite{noauthor_materials_nodate}.

\begin{table}
\caption{{\bf Error sources for the teleported \cnottxt.} The Bell-state fidelity with SPAM error contributions
subtracted is used as an estimate of the mixed-species \cnot fidelity \cite{noauthor_materials_nodate}.
$1\sigma$-uncertainty for the respective error sources are shown in parentheses.
}

\begin{center}
	\begin{tabular}{ll}
		\toprule
		 Source                  				      & Error ($10^{-2}$)   \\
		 \hline\hline
		 	 SPAM on two \beshort ions 	& 1.1(7) \\
 	 	SPAM on two \mgshort ions	    & 1.5(3) \\
		\mgone--\mgtwo Bell state			                      & 4.0(9)       \\
		\beone--\mgone \cnot           & 3.0(9)       \\
		\mgtwo--\betwo \cnot                  & 3(1)      \\
		 Coherence of \mgshort ions                            & 0.7(3)  	 \\
	 	Stray light from \mgone detection on \mgtwo 	  & 1.1(4) 		 \\
	 	Stray light from \betwo cooling on \beone         & 1.2(3) 		 \\

		 \midrule
		 Sum                                            &  16(2)  \\
		 Depolarizing model   						                  & 12(1)     \\
		 \bottomrule
	\end{tabular}
	\label{table:error_budget}
\end{center}
\end{table}

Ideally one would implement QGT using an ion species with a transition insensitive to magnetic field fluctuations to serve as both information-carrying qubits and entanglement-resource qubits, and a second dedicated species for cooling. This would mitigate decoherence and allow QGT to be embedded in a larger quantum circuit (in our experiment any prior information encoded in \beone and \betwo would be destroyed during cooling). Errors from stray light scattering could also be removed by using the coolant species for quantum logic readout \cite{schmidt_spectroscopy_2005}. To be viable for fault-tolerant error correction, larger algorithms like QGT will also require constituent operations to be performed with higher fidelity in multiple locations. A larger QCCD array would then have many different interaction zones and integrated detection zones \cite{slichter_uv-sensitive_2017}. The fact that our experimental duty cycle was dominated  by shuttling and associated recooling \cite{noauthor_materials_nodate} emphasizes the importance of cold diabatic transport \cite{bowler_coherent_2012,walther_controlling_2012} and faster cooling techniques \cite{roos_experimental_2000}.

Deterministic teleported \cnot gates
can serve as a useful primitive for large-scale quantum computation.
The integration of several operations, including mixed-species coherent control, ion transport, and entangling operations on selected subsets of qubits will be essential for building large-scale quantum computers based on ions in the QCCD architecture.
Moreover, applying consistency checks to the experimental data facilitated the identification of  error
sources in the experimental setup, illustrating the importance of performing
such checks in addition to tomography when characterizing quantum processes. Similar consistency checks could be done between disjunct processing nodes  executing the same routine, exposing compromised nodes that behave differently from the rest.

\bibliographystyle{Science}

\section*{Acknowledgments}
We thank P.~Hou and D.~Cole of the NIST Ion Storage Group for helpful comments on the manuscript.
We thank D. T. C. Allcock, S. C. Burd (both of the NIST Ion Storage Group),  the Oxford University, and the ETH Z\"urich ion trapping
groups for their advice on stabilizing magnetic fields.
{\bf Funding:} This work was supported by the Office of the Director of National Intelligence (ODNI) Intelligence
Advanced Research Projects Activity (IARPA), ONR, and the NIST Quantum Information Program.
D.K.\ acknowledges support from the Swiss National Science Foundation under grant no. 165208. S.D.E.\
acknowledges support by the U.S. National Science Foundation under Grant No. DGE 1650115.
Y.W., D.K., J.J.W., and H.M.V. are associates in the Professional Research Experience Program (PREP) operated jointly
by NIST and University of Colorado Boulder.
H.M.V. acknowledges support from the Schlumberger Foundation's Faculty for the Future program. {\bf Author contributions:} Y.W. and D.K. performed the majority of experiments with assistance from S.D.E., T.R.T, and J.J.W.. D.J.W., A.C.W., and D.L. advised on the project. K.H.M., H.M.V., S.G., and E.K. developed statistical models and provided theorectical support. Y.W., D.K., and K.H.M. contributed to data analysis. Y.W., S.D.E., and K.H.M. wrote the manuscript. All authors revised and commented on the manuscript.
{\bf Competing interests:} The authors declare that there are no competing financial interests.
{\bf Data and materials availability:} Data from the main text and supplementary materials are available through NIST public data repository \cite{wan_data_2019}.

\section*{List of Supplementary Materials}
Supplementary Text \\
Figure S1--S9 \\
Table S1--S3 \\
References \textit{(30--40)}

\clearpage

\begin{figure}[h]
	\centering
	\includegraphics[width=1.0\textwidth]{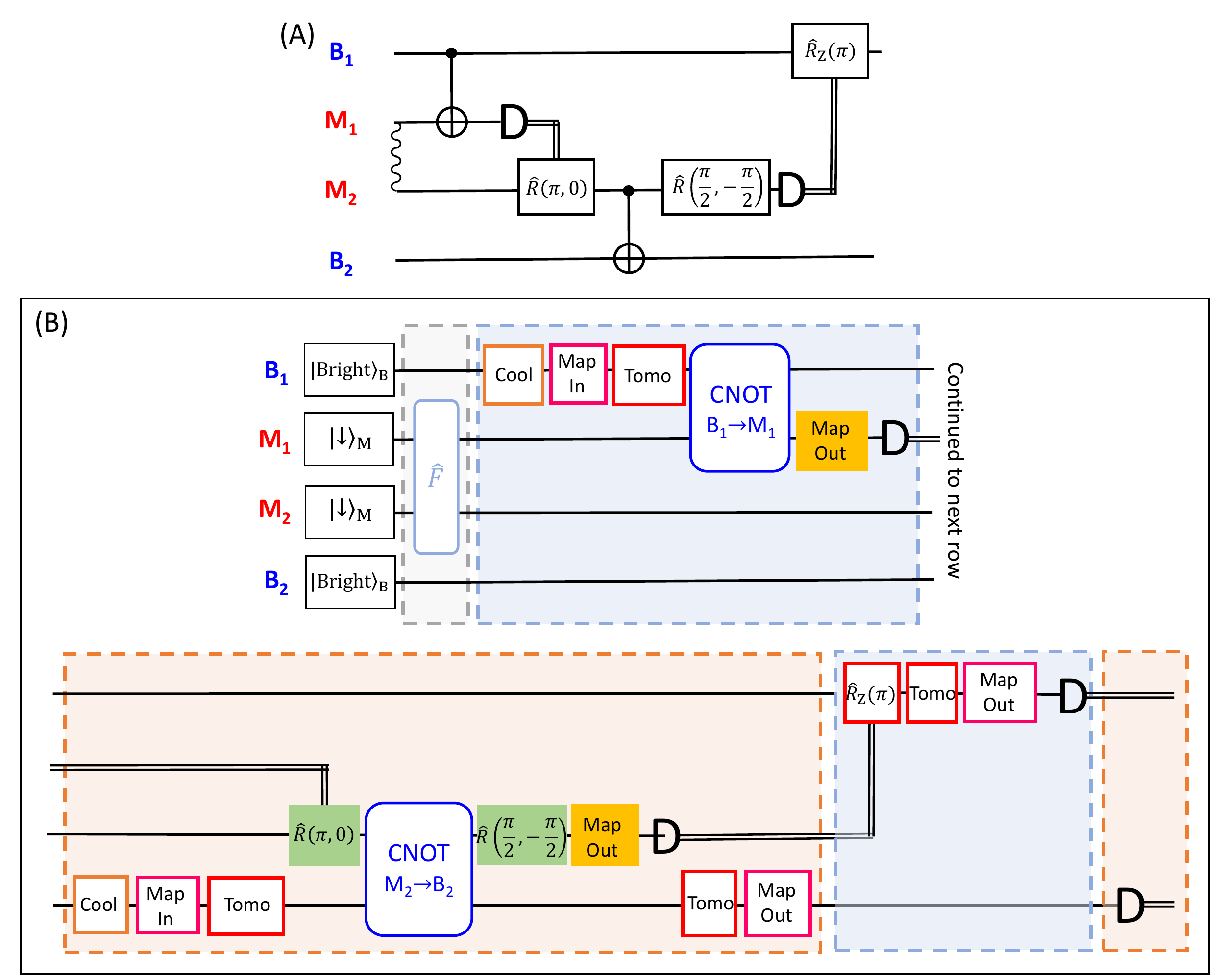}  
	\caption{\textbf{QGT circuit diagram.} (A)
		Circuit diagram for a teleported \cnot
		between qubits B$_\mathrm{1}$ and B$_\mathrm{2}$ as proposed in \cite{eisert_optimal_2000}.
		The wavy line represents entanglement, and double solid lines represent classical communication.
		(B) Experiment-specific circuit diagram for the teleported \cnot gate between \beone and  \betwo.
		``Map In" indicates mapping from the measurement basis to the computational basis, while ``Map Out'' indicates the opposite process. ``Tomo'' refers to single-qubit rotations for QPT.
	}
	\label{fig:fig1}  
\end{figure}

\begin{figure}[h]
	\centering
	\includegraphics[width=0.95\textwidth]{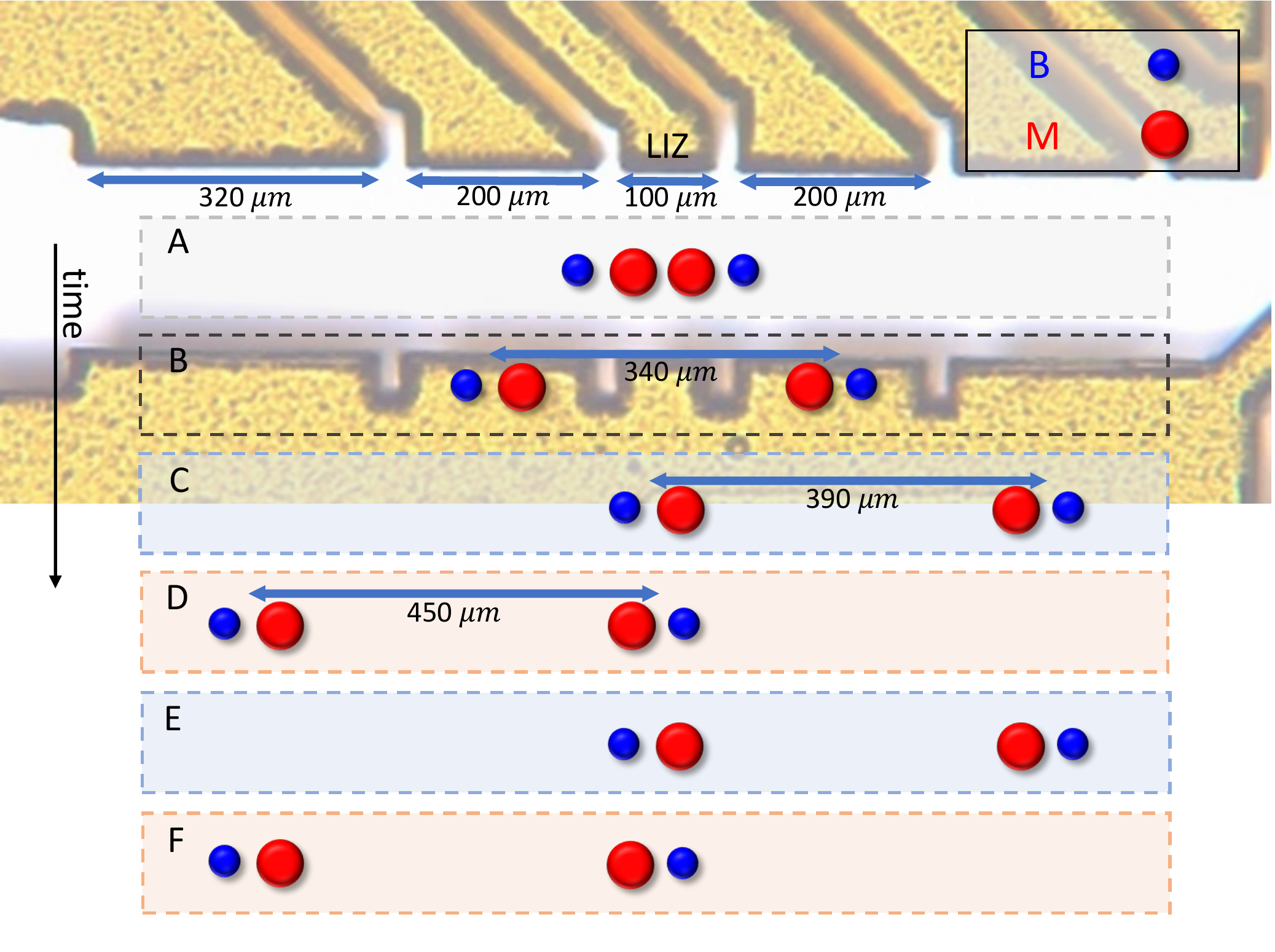}  
	\caption{\textbf{QGT shuttling sequence.}
		Panels~A-F show the shuttling sequence, overlaid on a photograph of a section of the trap electrode structure (ions and ion spacings are not to scale).
		After preparing the \mgshort ions in a Bell state to serve as the entanglement resource, the \bmmbnotation chain
		is split into two pairs of \bmpair ions, which are translated into and out of LIZ
		to address and detect individual ions (blocks C, D, E and F).
	}
	\label{fig:fig2}  
\end{figure}

\begin{figure}[h]
	\centering
	\includegraphics[width=1.0\textwidth]{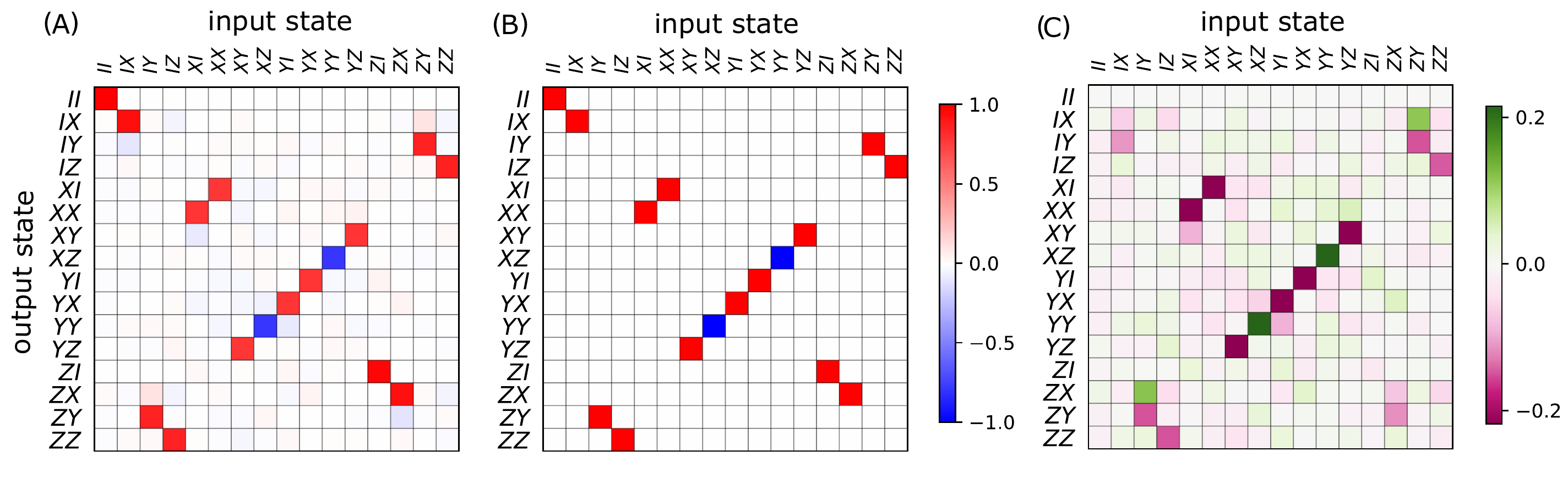}  
	\caption{\textbf{Pauli transfer matrix.} Visualization of quantum processes in the Pauli
		transfer matrix representation for
		(A) the experimental data ``Data Set 2'', (B) the ideal \cnot process, and (C) the difference between the experimental and  ideal process matrices \cite{noauthor_materials_nodate}.
		The Pauli transfer matrix of a process maps an arbitrary input density matrix, expressed as a linear combination of Pauli products, into the corresponding linear combination that describes the output density matrix.
		For our implementation, a 95\% confidence interval [0.845, 0.872] for the entanglement fidelity is determined with respect to an ideal \cnot.
	}
	\label{fig:fig3}  
\end{figure}

\clearpage
\newpage
\section*{Supplementary Text}

\renewcommand\thefigure{S\arabic{figure}}    
\setcounter{figure}{0}    

\renewcommand\thetable{S\arabic{table}}    
\setcounter{table}{0}

\begin{figure}[h]
	\centering
	\includegraphics[width=0.8\textwidth]{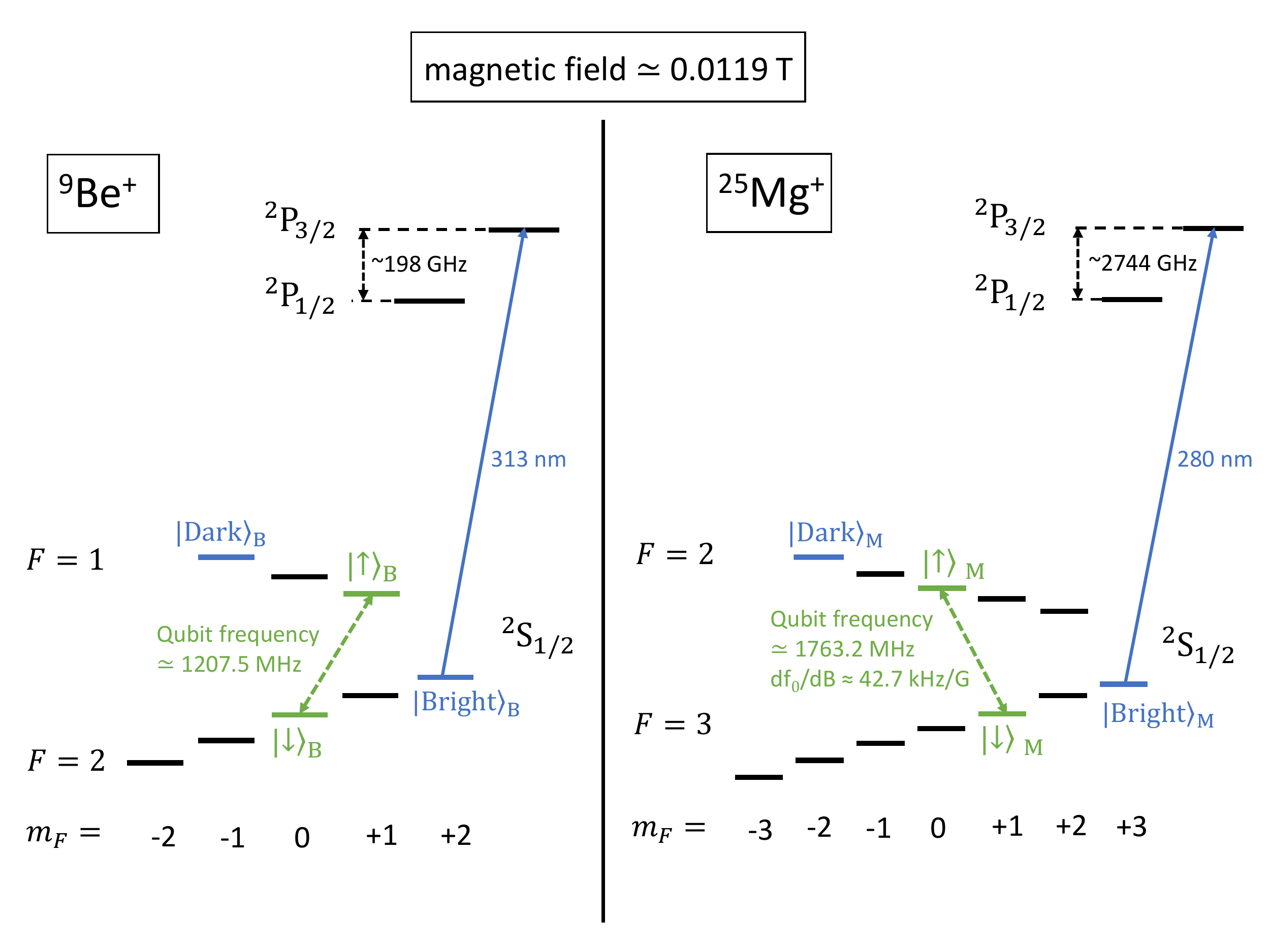}
	\caption{\textbf{Relevant level structures of beryllium and magnesium ions.}
		The \benine qubits are encoded in the computation basis $\ket{\uparrow}_\besubs$ and
		$\ket{\downarrow}_\besubs$ and mapped out to the measurement basis $\ket{\mathrm{Bright}}_\besubs$ and
		$\ket{\mathrm{Dark}}_\besubs$ respectively for fluorescence detection with $313~\mathrm{nm}$ light resonant with the $^\mathrm{2}$S$_\mathrm{1/2}$\,$\ket{\mathrm{Bright}}_\besubs$ $\leftrightarrow$ $^\mathrm{2}$P$_\mathrm{3/2}$\,$\ket{3, 3}_\besubs$ transition. The
		\mgfive qubits are encoded in the computation basis $\ket{\uparrow}_\mgsubs$ and $\ket{\downarrow}_\mgsubs$
		and mapped out to the measurement basis $\ket{\mathrm{Dark}}_\mgsubs$
		and $\ket{\mathrm{Bright}}_\mgsubs$ respectively for fluorescence detection with $280~\mathrm{nm}$ light 
		resonant with the 
		$^\mathrm{2}$S$_\mathrm{1/2}$\,$\ket{\mathrm{Bright}}_\mgsubs$ $\leftrightarrow$ $^\mathrm{2}$P$_\mathrm{3/2}$\,$\ket{4, 4}_\mgsubs$ transition. (See text) 
	}
	\label{fig:level_structure}
\end{figure}

\paragraph*{\paratrapandqubit}
This experiment is performed in a multi-layer segmented linear Paul trap, described in detail in \cite{blakestad_near-ground-state_2011}. We use \benine and \mgfive as qubits at a  magnetic field of ~$1.19\times 10^{-2}$~T. The detailed level structures for both qubits are depicted  in
Fig.~\ref{fig:level_structure}.
The qubits are encoded in the two levels of the hyperfine groundstate manifolds
$\ket{1,1}_\besubs \equiv \ustate_\besubs$ and $\ket{2, 0}_\besubs\equiv\dstate_\besubs$ for \benine ions and
$\ket{2, 0}_\mgsubs\equiv\ket{\uparrow}_\mgsubs$ and $\ket{3,
1}_\mgsubs\equiv\ket{\downarrow}_\mgsubs$ for \mgfive ions.
As in the main text, we use the symbol \beshort (\mgshort) for \benine (\mgfive) ions, and label all states of \benine (\mgfive) ions with the subscript \beshort (\mgshort)  for the remainder of the supplementary materials. 
At our chosen magnetic field, the \beshort qubits are first-order insensitive to magnetic field fluctuations. A Ramsey experiment with microwave pulses shows contrast higher than $90\%$ after
a Ramsey wait time of $1\,\mathrm{s}$; a Ramsey experiment with co-propagating laser carrier pulses shows similar results. 
The \mgshort qubits are first-order sensitive to magnetic field fluctuations with a linear coefficient of
$\sim\,$$430\,\mathrm{kHz\,mT}^{-1}$ \cite{tan_multi-element_2015}. 
We stabilize the magnetic field by measuring the current applied to the magnetic field coils 
and feeding back on the voltage setpoint of the current supply to keep the measured current stable. 
Individual components of the algorithm are generally calibrated in experiments triggered on the
$60\,\mathrm{Hz}$ AC line, but these components are typically used at different phases of the $60\,\mathrm{Hz}$ AC line when applied in calibration experiments and in the QGT algorithm. 
To mitigate miscalibrations from this, we modulate the magnetic field current feedback
setpoint with $60\,\mathrm{Hz}$ and $180\,\mathrm{Hz}$ signals  (phases relative to the AC cycle and amplitudes determined
experimentally) to reduce the magnetic field noise amplitude at these frequencies. 
We additionally synchronize the QGT algorithm with the $60\,\mathrm{Hz}$ AC line cycle
to reduce decoherence caused by the remaining magnetic field noise at $60\,\mathrm{Hz}$ and its
harmonics.
These improvements result in a coherence time of $\sim\,$$140\,\mathrm{ms}$ on the \mgshort qubit
transition, measured independently in a Ramsey experiment using microwave pulses and in another Ramsey experiment using co-propagating Raman beams.
Slow fluctuations in the ambient magnetic field noise lead to variations 
of the coherence time by up to $30\%$ from day to day.   

All laser beams used in this experiment 
are aligned to the ions at the laser interaction zone (LIZ), see Fig.~2 in the main text.
At this location in the trap, 
both \beshort and \mgshort ions experience micromotion along the direction the ions  are aligned in
(axial) with a modulation index of 
$2.67$ for \beshort ions 
and $1.08$ for \mgshort ions \cite{blakestad_high-fidelity_2009}.   This micromotion arises due to trap imperfections. 
An electro-optic modulator (EOM) is placed in one of the
\beshort motion-sensitive Raman beamlines to compensate for this effect, which results in
an effective modulation index of less than $0.1$ for \beshort; \mgshort axial micromotion is left uncompensated.
All \beshort motion-sensitive Raman transitions in this
work are driven on the micromotion carrier instead of the second micromotion sideband as
done in previous work in this trap  \cite{tan_multi-element_2015,gaebler_high-fidelity_2016}. 

\paragraph*{\paracp} Prior to running experiments, we load a \fouriontxt of two \beshort ions 
and two \mgshort ions into a \singlewell located at the LIZ with single-\beshort-ion trap frequencies of $2.0\,\mathrm{MHz}$ (axial), $12.0\,\mathrm{MHz}$
and $12.3\,\mathrm{MHz}$ (radial, two orthogonal transverse directions to the crystal axis).

The initial order of ions in the ion crystal is random and can change on  time-scales of minutes from background gas collisions.
To set the order before each shot of the experiment, we increase the axial
potential (to produce single-\beshort-ion trap frequencies of \tfinmhz{5.5} axial, \tfinmhz{10.5} and \tfinmhz{12.5} radial, but single-\mgshort-ion radial frequencies are considerably lower due to the inverse mass dependence of the radial pseudo-potential)
while simultaneously laser cooling both species.
In this case, the \fouriontxt is deformed into an energetically favorable diamond-shaped
configuration where the two \mgshort ions are aligned radially and displaced symmetrically about the $z$-axis between the two \beshort
ions.
Relaxing the trap back to the original confining potential brings the crystal 
to an ordered linear chain of \bmmbnotation, 
with trap frequencies of \tfinmhz{1.4}, \tfinmhz{3.0}, \tfinmhz{4.1}, and \tfinmhz{4.2} for the four
 axial normal modes. 

\paragraph*{\parashuttling} 
To individually address the two pairs of \bmpair ions, the \fouriontxt
is separated into \beone--\mgone and \mgtwo--\betwo in a process taking $570\,\us$
(Fig.~2A$\,\to\,$2B). The two ion pairs can then be separately transported
to the LIZ by applying shuttling steps from a set of four primitives, each taking $230\,\us$. This set includes shuttling from the ion positions shown in 
Fig.~2B$\,\to\,$2C, 2B$\,\to\,$2D, and their reversals.
The full shuttling sequence is composed from these basic separation and transport operations \cite{blakestad_high-fidelity_2009,blakestad_near-ground-state_2011}. While residing in the LIZ, the pairs of \bmpair ions are confined in a potential with trap frequencies of
\tfinmhz{2.1} and \tfinmhz{4.5} for the two axial normal modes. 

\paragraph*{\paraspam}
To prepare for each experiment, the \beshort and \mgshort ions are optically pumped to \bebrstate and \mgbrstate respectively. The \mgshort ions are then
transferred to the state  $\ket{\downarrow}_\mgsubs$ with microwave pulses.
To detect their internal states, the ions are mapped from the computation basis to the measurement
basis. The \beshort ions are mapped individually with co-propagating Raman beams from
$\ustate_\besubs$ to \bebrstate and from $\dstate_\besubs$ to \bedkstate, while
the \mgshort ions are mapped with a combined scheme 
\begin{align}
	 \dstate_\mgsubs &\xrightarrow{\text{laser}}  \ket{2, +2}_\mgsubs 
	 \xrightarrow{\text{microwave}}
	 \mgbrstate
	 \\
	 \ustate_\mgsubs &\xrightarrow{\text{laser}}  \ket{3, -1}_\mgsubs
	 \xrightarrow{\text{microwave}}
	 \mgdkstate
\end{align}
to maintain both individual addressing and high transfer efficiency\footnote{In contrast, all
qubit operations for both species are implemented with laser pulses.}.
After mapping out to the measurement basis, the internal states of the ions are distinguished by fluorescence detection for a duration of $330\,\us$
for \beshort and $200\,\us$ for \mgshort ions. An EOM is placed in the \beshort resonant beamline to 
approximately cancel the effect of axial micromotion. We obtain an approximately Poissonian
distribution of photon counts  with a mean of roughly $30$ counts for each individual \beshort
and \mgshort ion in the bright state, and average background counts of $1.5$ and $0.8$
for \beshort and \mgshort respectively in the dark state. For conditional operations based on the \mgshort
measurements, a threshold of $10$ photon counts is set to distinguish $\ket{\mathrm{Bright}}_\mgsubs$ from $\ket{\mathrm{Dark}}_\mgsubs$ in a single
measurement. Thresholds for the \beshort measurements are discussed in the quantum process tomography section below.

\paragraph*{\parasbc}
At the beginning of the algorithm, we apply
Doppler cooling (DC) on first \mgshort and then \beshort ions followed by sideband cooling (SBC) on the \beshort
ions \cite{monroe_resolved-sideband_1995}.  
The SBC sequence consists of continuous SBC
applied sequentially to the four axial modes \cite{wan_efficient_2015} followed by a short sequence of pulsed SBC
\cite{monroe_resolved-sideband_1995}, which cools the \fouriontxt to an average 
motional quantum number \nbar of less than $0.1$ for the mode used to implement the \msgate gate and less than $0.3$ for all the
other axial modes. 
After separating the two \bmpair pairs and shuttling one pair of \bmpair ions to the LIZ (Fig~2C and 2D),
Doppler cooling is only applied to the \beshort ions to preserve the entanglement of the \mgshort
qubits. The
SBC on \beshort ions cools both axial modes of the \bmpair pairs
to an average occupation number \nbar of less than $0.1$
before applying two-qubit gate operations.

\paragraph*{\paragates}
Two sets of Raman laser beams are used to implement the single-qubit and two-qubit
operations in this experiment. For the \mgshort ions, a frequency-quadrupled diode laser
($\sim\,$$295\,\mathrm{GHz}$ blue-detuned from the S$_\mathrm{1/2}$ $\leftrightarrow$
P$_\mathrm{3/2}$ transition) is used to produce all the frequencies necessary
for driving co-propagating carrier transitions, motion-sensitive carrier transitions, 
and two-qubit \msfull (\msgate) gates. 
Similarly, all laser beams for addressing the \beshort ions are derived from a fiber laser system
($\sim\,$$265\,\mathrm{GHz}$ red-detuned from the S$_\mathrm{1/2}$ $\leftrightarrow$
P$_\mathrm{1/2}$ transition). The co-propagating carrier pulses induce single-qubit rotations
\begin{equation}
	\hat{R}(\theta, \phi) = \begin{bmatrix}
		\cos{\frac{\theta}{2}} & -i\mathrm{e}^{-i\phi}\sin{\frac{\theta}{2}} \\
 		-i\mathrm{e}^{i\phi}\sin{\frac{\theta}{2}} &  \cos{\frac{\theta}{2}}
	\end{bmatrix}
\end{equation}
where $\theta$ is the angle rotated and $\phi$ is the angle between the rotation axis
(in the $xy$ plane) and the positive $x$-axis of the Bloch sphere.

We use a pair of \mgshort ions in the Bell state $\phiplusmg=\bellstaters$
as the entanglement resource for the teleported \cnot gate.
This entangled pair is generated with an \msgate interaction applied only to the \mgshort ions 
in the \fouriontxt.
The gate is implemented on the out-of-phase
mode at \tfinmhz{3.0} with a gate duration of about $56\,\us$. 
The \msgate interaction implemented here does not generate a deterministic Bell state due to a slowly-fluctuating interferometric phase between the two arms of the motion-sensitive Raman beams.

\begin{figure}[h]
	\centering
	\includegraphics[width=0.7\textwidth]{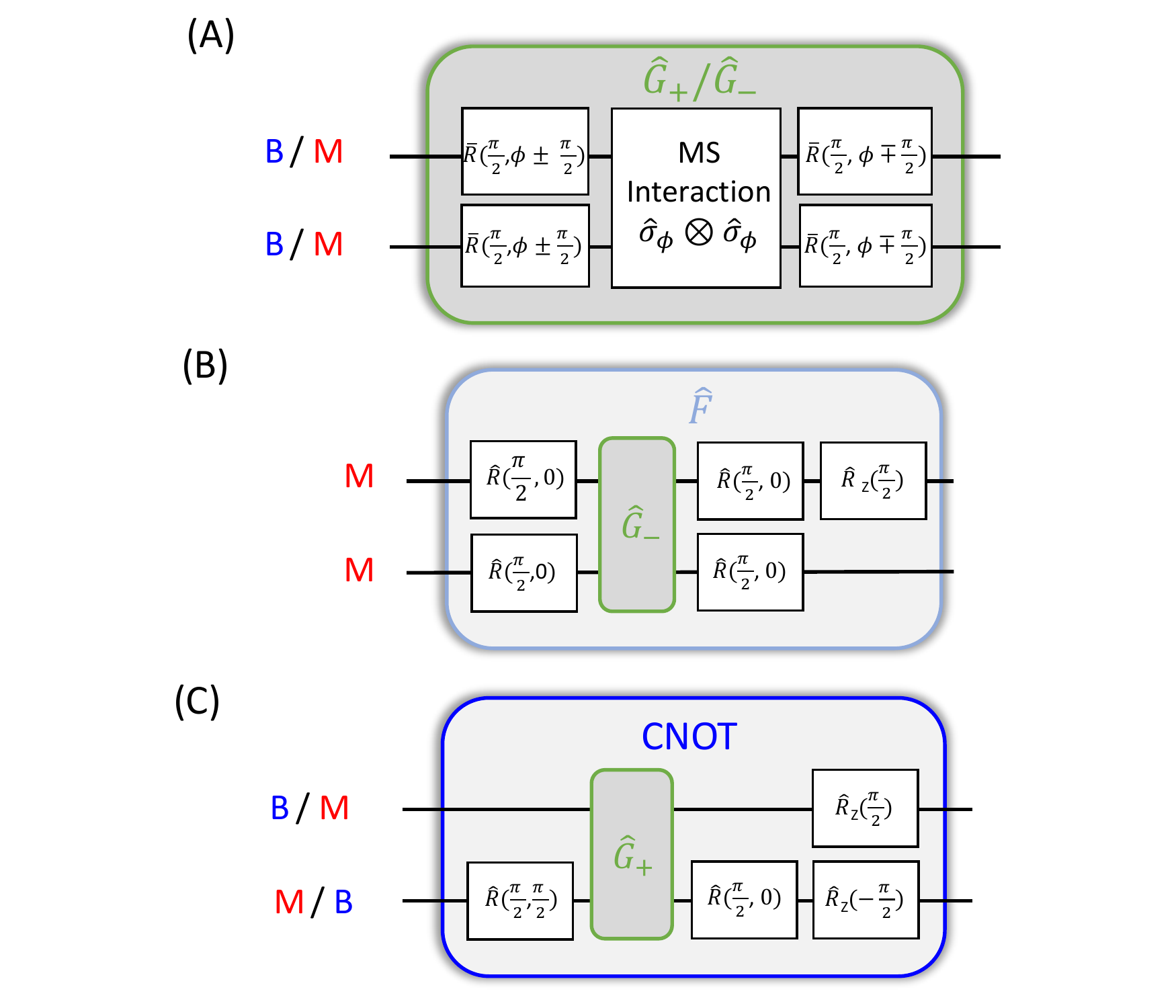}
	\caption{{\bf Composite gates.} (A) Phase gates $\hat{G}_\mathrm{\pm}$ are constructed by surrounding the \msgate pulse with single-qubit rotations $\bar{R}$,
		all driven by motion-sensitive Raman beams. The notation $\bar{R}$ distinguishes  itself from motion-insensitive single-qubit rotations $\hat{R}$ driven by co-propagating Raman beams.
		The resulting operation does not depend on the uncontrolled interferometric phase of
		the Raman beam sets as long as this phase is constant during the entire $\hat{G}_\mathrm{\pm}$ gate implementation. The sign of the implemented phase gate depends on the sign of \msgate detuning from motional sidebands, phases of surrounding single-qubit rotations, and motion phases in the case of mixed-species gates.
		(B)	The \mgone--\mgtwo Bell-state-generating gate $\hat{F}$ is constructed by surrounding $\hat{G}_\mathrm{-}$ with co-propagating $\pi/2$ pulses and phase-shifting one of the qubits by $\pi/2$ at the end. The  sequence generates the \mgone--\mgtwo Bell state $\phiplusmg = \bellstaters$ from the $\ket{\downarrow \downarrow}_\mgsubs$ state.
		(C) The \cnot gate is constructed by surrounding $\hat{G}_\mathrm{+}$ with co-propagating $\pi/2$ pulses on the
		target qubit and phase shifting the control and the target qubit by $\pi/2$ and $-\pi/2$ respectively.}
	\label{fig:composite_gates}
\end{figure}

To mitigate this, the \msgate pulse is surrounded with a
pair of carrier $\pi/2$ pulses using the same set of motion-sensitive laser beams, as shown in Fig.~\ref{fig:composite_gates}A, resulting in phase gates 
\begin{equation}
\hat{G}_\mathrm{+}=\begin{bmatrix}
1       & 0 & 0  & 0 \\
0       & i & 0  & 0 \\
0   & 0 & i  & 0 \\
0    & 0 & 0  & 1 
\end{bmatrix}
\;\text{or}\;\;
\hat{G}_\mathrm{-}=\begin{bmatrix}
1   & 0  & 0  & 0 \\
0   & -i & 0  & 0 \\
0   & 0  & -i & 0 \\
0    & 0  & 0  & 1 
\end{bmatrix},
\end{equation}
that do not depend on the fluctuating interferometric phase \cite{lee_phase_2005,tan_multi-element_2015}. 
The sign of the implemented phase gate depends on the sign of the \msgate detuning from the motional sidebands, phases of the surrounding single-qubit rotations, and in the case of mixed-species gates on the motion phases which define the direction of the ions' displacement in phase space \cite{lee_phase_2005,tan_multi-element_2015}. 

When constructing our \mgone--\mgtwo Bell state, the \msgate detuning was chosen to be positive, and the two \mgshort ions share common single-qubit-rotation phases from  global laser beams. This results in the phase gate $\hat{G}_\mathrm{-}$, which is further surrounded with a pair of co-propagating carrier $\pi/2$ pulses to construct the Bell-state-generating gate $\hat{F}$
used to entangle the two \mgshort ions (Fig.~\ref{fig:composite_gates}B). We further verify through a parity scan that the \mgone--\mgtwo Bell state is $\phiplusmg = \bellstaters$ after applying $\hat{F}$ to $\ket{\downarrow\downarrow}_\mgsubs$.

The \cnot gates within the \bmpair pairs are implemented with
a mixed-species \msgate interaction on the in-phase mode at \tfinmhz{2.1} \cite{tan_multi-element_2015}.
We produce a phase gate similarly as done for $\hat{F}$, but with two pairs of perpendicular Raman beams, each pair addressing only one species. This gives additional degrees of freedom in the form of a differential motion phase between species and different phases for the surrounding single-qubit rotations, resulting in ambiguity about the resulting phase gate. 
To ensure that we produce the right form of phase gate ($\hat{G}_\mathrm{+}$), we calibrate the phases of the single-qubit rotations by running a complete mixed-species CNOT gate (Fig.~~\ref{fig:composite_gates}C) while looking for the appropriately conditioned bit flip on the target.
Here, the $\mathrm{CNOT}_\mathrm{\beone\rightarrow\mgone}$ 
($\mathrm{CNOT}_\mathrm{\mgtwo\rightarrow\betwo}$) with \beshort (\mgshort) as the control is constructed  by surrounding the phase gate $\hat{G}_\mathrm{+}$ with additional co-propagating carrier $\pi/2$ pulses on the \mgshort (\beshort) ion.

The rotation 
\begin{equation}
	\hat{R}_\mathrm{Z}(\alpha) = \begin{bmatrix}
	\mathrm{e}^{-i\alpha/2} & 0 \\
	0 &  \mathrm{e}^{i\alpha/2}
	\end{bmatrix}
\end{equation}
at the end of the pulse groups is implemented in software by changing the phase of subsequent single-qubit rotations for that qubit by $-\alpha$. 
This is possible because the two-qubit phase gates $\hat{G}_\mathrm{\pm}$ commute with $\hat{R}_\mathrm{Z}(\alpha)$ rotations.


\paragraph*{\parapt}
The experiment is performed with a direct digital synthesizer (DDS) running in the ``absolute
phase" mode, where the phase of the DDS is reset if a frequency or phase change is necessary. The phase evolution of the qubits in the
laboratory frame is tracked on the experiment control computer by accounting for the free
precession frequency $f_\mathrm{0}$ of the qubits and the precise timing $t$ from the first pulse to
the current pulse. To apply a rotation at phase $\phi$, the corresponding DDS is set to the phase
$2\pi\cdot f_\mathrm{0}\cdot t + \phi + \phi_\mathrm{0}$, where the additional offset
$\phi_\mathrm{0}$ takes into account the phase shifts induced by the AC-Stark effect and magnetic
field gradient. This removes the requirement of tracking the phase of each individual qubit with an
independent DDS. In this experiment, we calibrate the phase offset $\phi_\mathrm{0}$ for each
co-propagating carrier pulse with a separate Ramsey experiment. 
This could be avoided in future experiments by calibrating the AC-Stark shift 
and the magnetic-field-gradient-induced
phase shift and calculating the required phases from these calibrations.

\begin{figure}[h]
	\centering
	\includegraphics[width=0.8\textwidth]{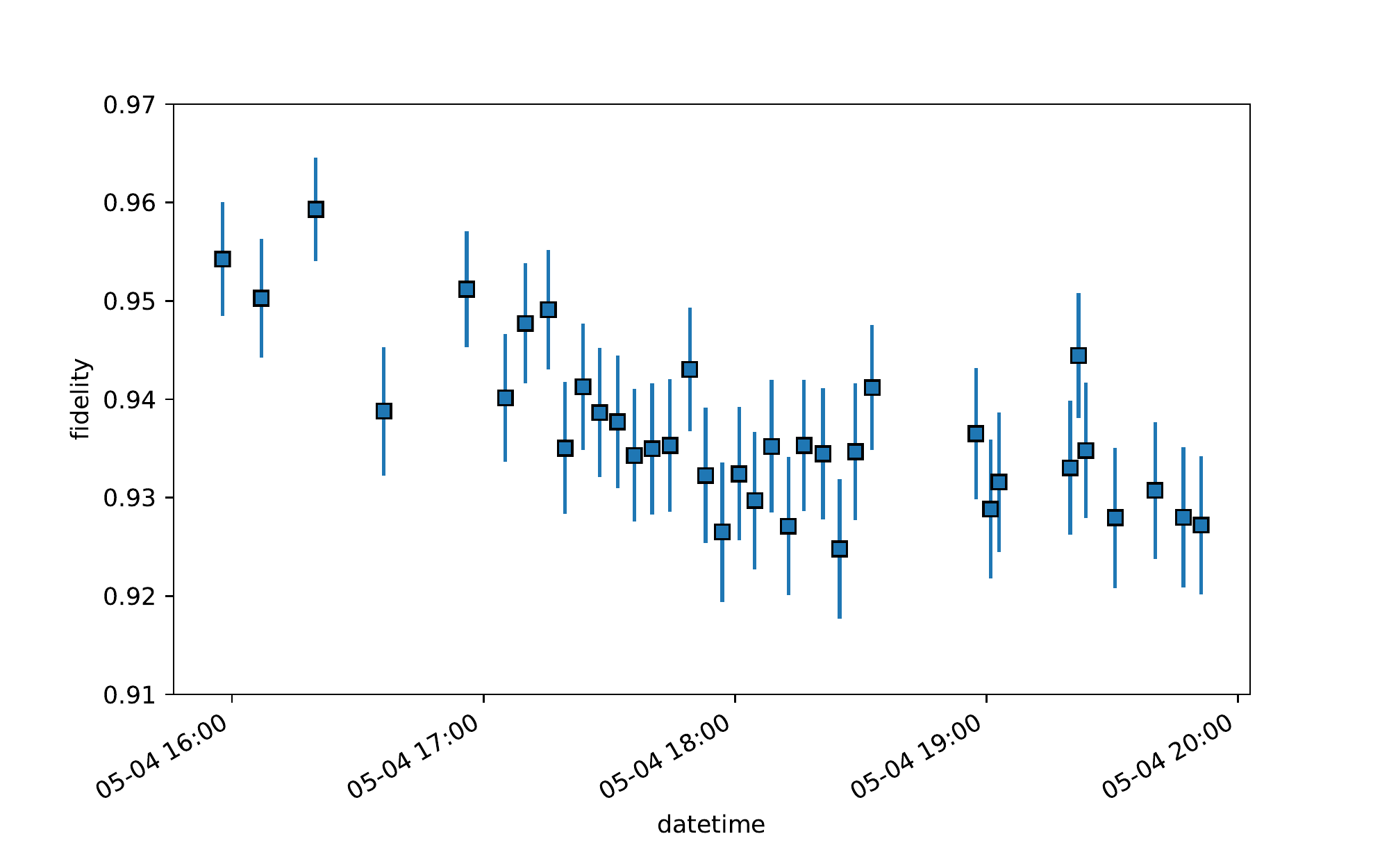} 
	\caption{\textbf{Drifts of gate fidelity.} The fidelity of the \mgone--\mgtwo Bell state
		generated by $\hat{F}$ was measured continuously over a period of four hours without
		re-calibration of experimental  parameters, showing the drift away from
		the initial value over time. The exact direction and rate of the drift varies from run to run. 
		The data shown here was taken on May~04,~2018. The gaps near 16:00 and 18:30 are caused by  reloading the ions.
	} 
	\label{fig:drifts}
\end{figure}

\paragraph*{\paradc}
To improve the long-term stability of our experiment, parameters such as pulse lengths and
transition frequencies are re-calibrated during data acquisition to reduce the effect of slow drifts in laser
output, beam pointing, and magnetic field strength. We observe that these experimental parameters
 drift significantly over time-scales of several minutes to hours, which affects the fidelity of the algorithm and all its components.
For example, Fig.~\ref{fig:drifts} shows the measured fidelity of the \mgone--\mgtwo Bell state
(without correcting for state preparation and measurement (SPAM) errors) over a period of four hours without recalibration.
The fidelity of the Bell state drifts away from the value obtained after calibration.
Therefore, to maintain all of the operations used at a nearly constant and high fidelity, running
the algorithm is conditioned upon fulfilling several validation experiments (validators). 
The validators verify that parameters were sufficiently
accurate within a predetermined time period before the algorithm is scheduled, otherwise
re-calibration is triggered.
Each validator is in turn conditioned on its dependencies. 
Consider the \beshort co-propagating carrier rotation $\hat{R}(\pi/2, 0)$ as an example: with \beshort ions initialized
in $\ustate_\besubs$, the carrier $\pi/2$ pulses are validated by observing that 1) the probability of four successive pulses bringing
the ions back to $\ustate_\besubs$ and 2) the probability of six successive pulses bringing the ions
to $\dstate_\besubs$ are both higher than their pre-defined thresholds. Failing the validation will trigger recalibration of the
pulse parameters (pulse duration, transition frequency, and phase offset between subsequent
pulses) and validators of the pulse parameters' dependencies. Here, the calibration of the pulse parameters of the \beshort co-propagating carrier rotations relies on the correct values of magnetic field and mapping pulses, which serve as the dependencies.

\begin{figure}[h]
	\centering
	\includegraphics[width=0.9\textwidth]{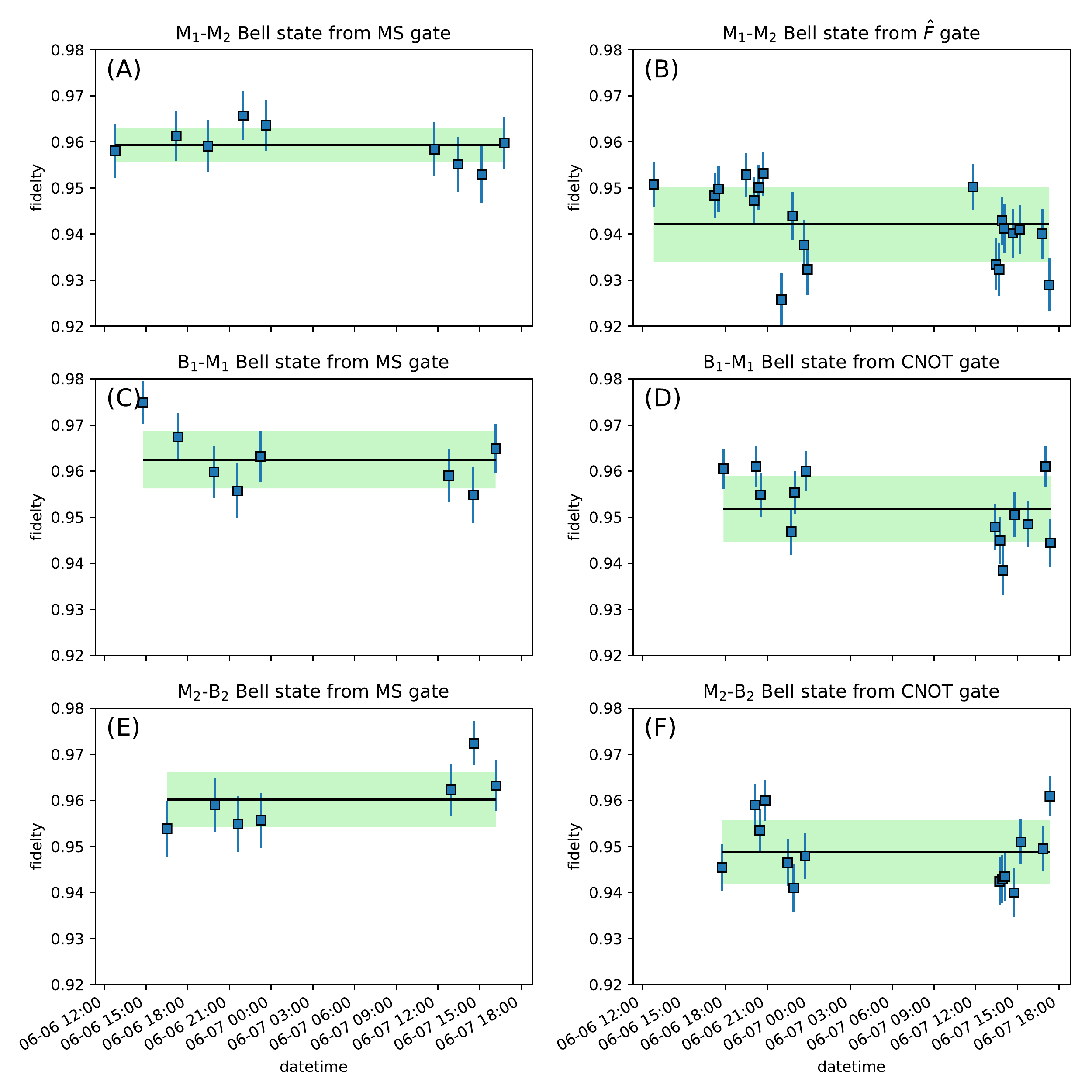}
	\caption{\textbf{Bell-state fidelities.} Bell-state fidelity measurements interspersed while taking full process tomography data on the teleported \cnot, without correcting for SPAM. The measurement spans over two days from June~6 to June~7,~2018.
		(A) \mgone--\mgtwo Bell state from the \msgate gate.  
		(B) \mgone--\mgtwo Bell state from $\hat{F}$. For (A) and (B), microwave pulses are used
		for mapping into and out of the computation basis.
		(C) \beone--\mgone Bell state from \msgate gate. (D) \beone--\mgone Bell state
		from \cnot gate. (E) \mgtwo--\betwo Bell state from \msgate gate. (F)
		\mgtwo--\betwo Bell state from \cnot gate. Squares: fidelity data. Solid black
		line: mean fidelity. Green band: standard deviation of the data points.
		The error bars on the fidelity data are the uncertainties of each fidelity measurement due to
		projection noise.    
	}  
	\label{fig:all_fidelity}
\end{figure}

Maintaining the fidelity of these operations at a consistent level significantly
mitigates the effect of experimental drifts. As shown in Fig.~\ref{fig:all_fidelity}, we track the fidelities of Bell states
generated from the \msgate, $\hat{F}$ and \cnot gates while taking tomography
data. 
The standard deviation of these measurements does not differ significantly from the 
uncertainty of the individual data points due to projection noise.

\paragraph*{\paratomography}
For the required informationally-complete set of measurements, the two \beshort ions were prepared in the following 16 different combinations of input states by single-qubit rotations after initial state preparation to the $\ket{\uparrow\uparrow}_\besubs$ state:

\begin{center}
	\begin{tabular}{llll}
		$\ket{ \uparrow \uparrow }_\besubs$   & $\ket{ \uparrow \downarrow }_\besubs$   & $\ket{ \uparrow \p}_\besubs$   &
		$\ket{\uparrow \r }_\besubs$	   \\
		$\ket{ \downarrow \uparrow }_\besubs$ & $\ket{ \downarrow \downarrow }_\besubs$ & $\ket{ \downarrow \p }_\besubs$ &
		$\ket{\downarrow \r }_\besubs$  \\
		$\ket{\p\uparrow }_\besubs$          & $\ket{ \p \downarrow }_\besubs$          & $\ket{ \p\p }_\besubs$          & $\ket{ \p\r }_\besubs$           \\
		$\ket{ \r \uparrow }_\besubs$          & $\ket{ \r \downarrow }_\besubs$          & $\ket{ \r\p }_\besubs$          &$\ket{\r\r }_\besubs$
	\end{tabular}
	
\end{center} 
where $\ket{\p}_\besubs = \tfrac{1}{\sqrt{2}}(\ket{\uparrow}_\besubs+\ket{\downarrow}_\besubs)$ and $\ket{\r}_\besubs =
\tfrac{1}{\sqrt{2}}(\ket{\uparrow}_\besubs + i\ket{\downarrow}_\besubs)$. For each of these 16 inputs, the output
states were measured along  nine different combinations of measurement axes
$\mathrm{XX}$, $\mathrm{XY}$, $\mathrm{XZ}$, $\mathrm{YX}$, $\mathrm{YY}$, $\mathrm{YZ}$,
$\mathrm{ZX}$, $\mathrm{ZY}$, $\mathrm{ZZ}$, where a measurement along the $\mathrm{Z}$ axis was
implemented by transferring to the measurement basis and performing fluorescence detection, and a measurement
along the $\mathrm{X}$ or $\mathrm{Y}$ axis was performed similarly after a rotation $\hat{R}(\pi/2, -\pi/2)$ or $\hat{R}(\pi/2, 0)$,
respectively.

\begin{figure}[h]
	\centering
	\includegraphics[width = 1.0\textwidth]{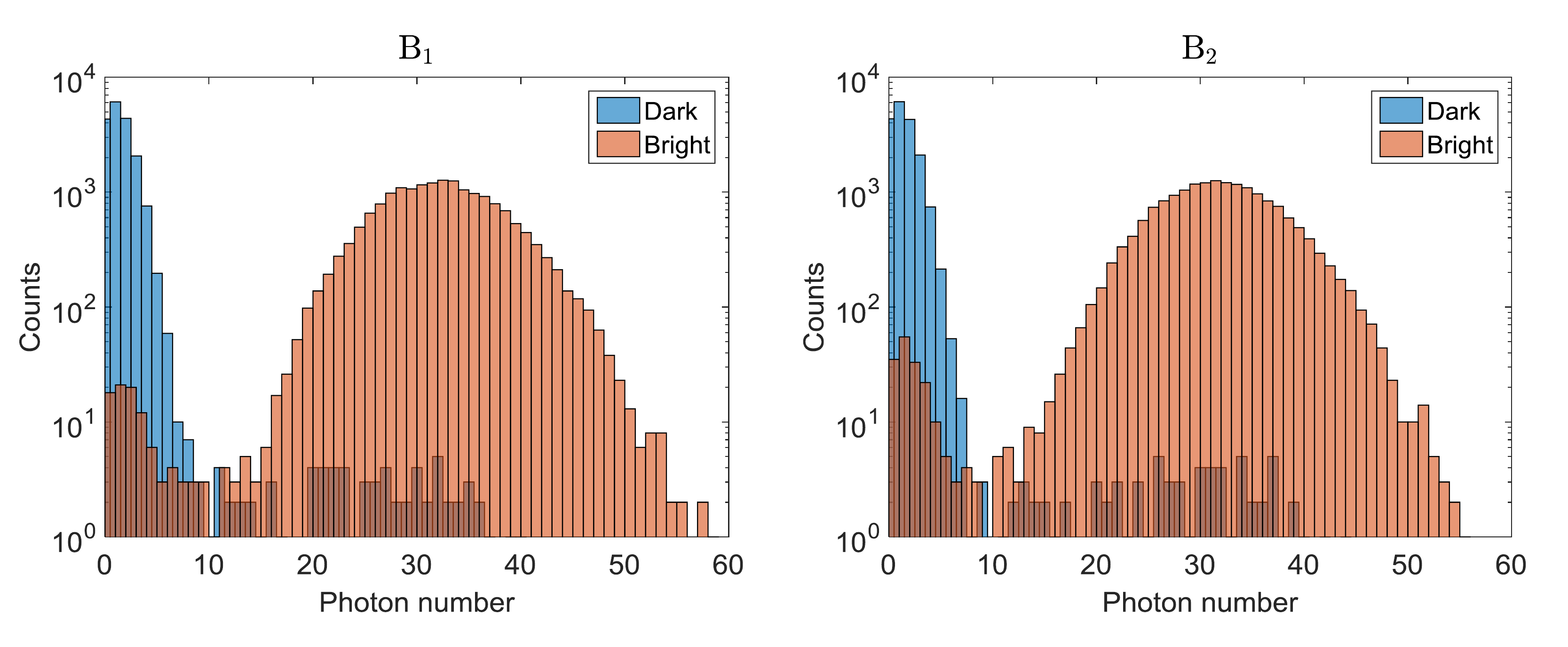}   
	\caption{\textbf{Reference histograms.} Histograms of photon number counts when \beshort ions are prepared in dark
		and bright states. Histograms are obtained from measurements taken during data acquisition. Photon count
		thresholds of 8 and 9 for \beone and \betwo, respectively, are determined to minimize the probability of misclassification. The
		reference histograms are used to infer the single-qubit POVM element $E_+$ corresponding to observing a photon
		number greater than the threshold: $E_+ =
		(1-p)\ket{\mathrm{Bright}}_\besubs\prescript{}{\besubs}{\bra{\mathrm{Bright}}}+p\ket{\mathrm{Dark}}_\besubs\prescript{}{\besubs}{\bra{\mathrm{Dark}}}$.
		We obtain $p = 0.0090$ and $p = 0.0134$ for $\beone$ and $\betwo$, respectively.
	}
	\label{fig:ref_hists}  
\end{figure}

We collected two full data sets for process tomography, and used dataset 1 to develop the final analysis
protocol which was then applied without further modifications to dataset 2.
Process tomography is performed by preparing the informationally-complete set of input states $\{\rho_k\}_{k=1}^{16}$ listed above,
applying the QGT algorithm to each input state, and measuring the two \beshort ions in one of the 9
possible Pauli-product bases. 
Each Pauli-product basis constitutes a positive operator-valued measure (POVM) with 4 elements, corresponding
to the detections of bright or dark states of the two ions. 
The family of 4$\times$9 POVM elements $\{E_l\}_{l=1}^{36}$ obtained from the
POVMs for the 9 Pauli-product bases is informationally complete for
state tomography.
In this section, we refer to running the algorithm with a single choice of state preparation and measurement basis as
an \textit{experiment}, and a single instance of input state preparation, QGT algorithm, and measurement as a \textit{trial}.
We assume perfect preparation of the input \beshort states, but to account for the small overlap between photon-count distributions of the bright and dark states during detection, we model the POVM elements
for individual ion measurements as a convex sum of projectors onto the bright and dark states. From a set of
reference photon count histograms recorded periodically between experiments during the data acquisition,
we infer the POVM elements by finding the thresholds and weights in the convex sum that maximize the likelihood
of the observed reference histogram data (See Fig.~\ref{fig:ref_hists}). This method for inferring the POVM elements is a special case of the measurement tomography procedure described in \cite{keith_joint_2018}.

Given a quantum process $\cE$ applied to input state $\rho_k$, the probability
of observing measurement outcome $E_l$ on a single trial is $\mathrm{Pr}(E_l | \rho_k)=\Tr(\cE(\rho_k)E_l)$. The probability of observing
all the recorded experimental data is given by the likelihood function:
\begin{equation}
L(\cE) = \prod_{kl}\Tr(\cE(\rho_k)E_l)^{n_{kl}}
\end{equation}
where $n_{kl}$ is the number of times outcome $E_l$ was observed when state $\rho_k$ was prepared. We estimate the process
using the method of maximum likelihood (ML), which involves maximizing the log-likelihood function $\cL(\cE) = \ln({L(\cE)})$
over all two-qubit completely-positive trace-preserving (CPTP) maps $\cE$.
We use the Choi matrix representation \cite{choi_completely_1975}, in which the process $\cE$ is
represented by the $d^2$-by-$d^2$ density matrix $\chi = (I\otimes\cE)(\ket{\Phi^+}\bra{\Phi^+})$, where $d=4$ is the dimension of the Hilbert space,  $\ket{\Phi^+}=\frac{1}{\sqrt{d}}\sum_{i=0}^{d-1}\ket{i}\ket{i}$
is a maximally entangled state between two copies of the Hilbert space,
 and $I$ is the identity matrix. In terms of the Choi matrix,
the action of the process on an input state is given by $\cE(\rho_k)=d\,\Tr_1\big(\chi(\rho^{\intercal}_k \otimes I)\big)$,
where $\Tr_1$ denotes a partial trace over the first subsystem and $^{\intercal}$ denotes transposition.
The log-likelihood function is then given by
\begin{equation}\label{eq: loglike}
\cL(\cE) = \sum_{kl}n_{kl}\ln\Big(d\,\Tr\big(\chi(\rho^{\intercal}_k \otimes E_l)\big)\Big)
\end{equation}
We use the ``$R\rho R$'' algorithm for processes \cite{Fiurasek_maximum-likelihood_2001} to find the Choi matrix $\hat{\chi}$
that maximizes the log-likelihood function. Throughout this section,
a ``hat'' placed above the symbol for a physical quantity is used to denote a statistical estimate of that quantity.
The \textit{entanglement fidelity} \cite{schumacher_sending_1996} with respect to $U=\cnot$ is defined to be 
\begin{equation}\label{eq: entanglement fidelity}
F(\cE,U)=\bra{\Phi^+}(I\otimes U^{\dag})\chi(I\otimes U)\ket{\Phi^+}
\end{equation} 
From the ML-estimated process we obtain the ML entanglement fidelities $\hat{F}_{ML} = 0.858$ for dataset 1 and $\hat{F}_{ML} = 0.851$ for dataset 2.  

\begin{figure}[h]
	\centering
	\includegraphics[width = 0.8\textwidth]{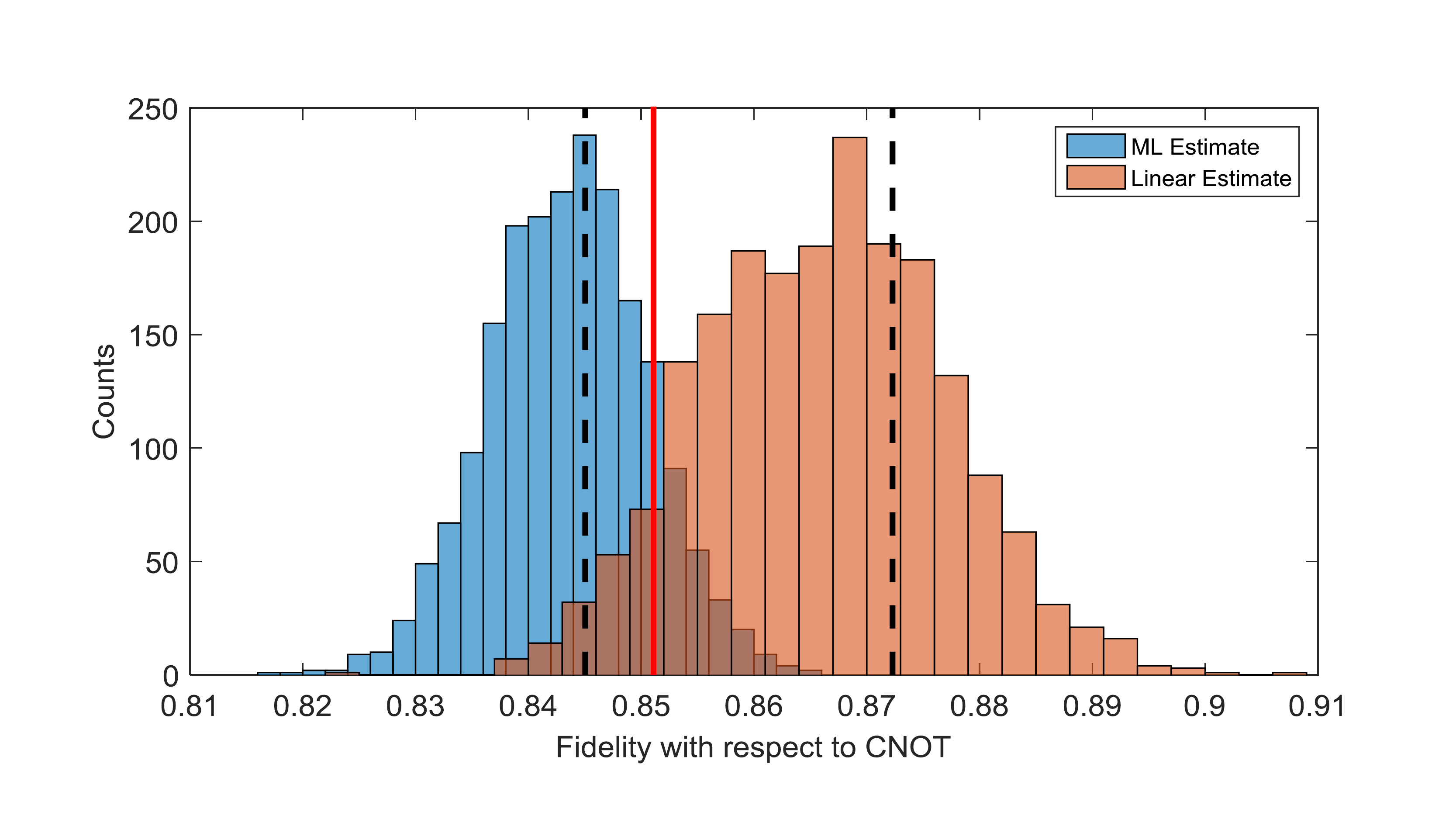}
	\caption{\textbf{Histograms of bootstrapped fidelity distributions.} The blue (left) histogram displays the distribution
		of process fidelities obtained from running ML on 2000 parametric bootstrapped data sets generated from the ML estimated process $\hat{\chi}$ of dataset 2.
		The red solid vertical line indicates the process fidelity of $\hat{\chi}$. The black vertical dashed lines mark the boundary of a
		95\% confidence interval computed via the basic bootstrap method. 
		The orange (right) histogram displays the distribution of fidelities obtained from applying the linear fidelity estimator $\hat{F}_L$ to 2000 non-parametric resamples of the experimental data set.
	}
	\label{fig:boot_fids}
\end{figure}

We obtain confidence intervals for the process fidelity using a parametric bootstrap method.
Associated with each confidence interval is a confidence level.  A confidence interval is defined as follows: in an ensemble of identically analyzed data sets (following the assumed model), one expects the frequency, with which the true value of the fidelity for any given data set will lie within  that data set's confidence interval, to equal the confidence level.
The ML estimate $\hat{\chi}$ is used to simulate 2000 synthetic data sets, and ML is run on each synthetic data set, producing a distribution
of bootstrapped fidelities shown in Fig.~\ref{fig:boot_fids}. We then compute the basic bootstrap confidence interval \cite{efron_introduction_1994},
which is defined as follows.
Let $\hat{F}_{ML}$ be the ML estimated fidelity, and for $\alpha\in[0,1]$, let $f_\alpha$ be the fidelity value
corresponding to the $(100\,\alpha)^{\mathrm{th}}$ percentile of the bootstrapped distribution.
Then the $100(1-2\alpha)$ percent confidence interval is $[2\hat{F}_{ML}-f_{(1-\alpha)},\quad 2\hat{F}_{ML}-f_{\alpha}]$.
The endpoints of the confidence interval are obtained by
reflecting the upper and lower percentile values of the bootstrapped distribution about the ML estimate.
This results in a confidence interval that approximately corrects for bias in the ML estimate (see Fig.~\ref{fig:boot_fids}).
Using this method, we obtain 95\% confidence intervals of $[0.852, 0.878]$ for dataset 1 and $[0.845, 0.872]$ for dataset 2. Confidence intervals for other confidence levels are in Table~\ref{tab:analysis_sum}.

As a consistency check against the ML fidelity estimate, we also constructed a linear estimator $\hat{F}_L$ that computes
the fidelity directly from the observed frequencies: $\hat{F}_L=\sum_{kl}a_{kl}f_{kl}$, where $f_{kl}$ is the observed frequency
of seeing outcome $E_l$ given state preparation $\rho_k$. To derive the coefficients $a_{kl}$, we first rewrite Eq.~\ref{eq: entanglement fidelity} as
\begin{equation}\label{eq: entanglement fidelity 2}
    F(\cE,U)=\frac{1}{d^2}\sum_{i,j=0}^{d-1}\bra{i}U^{\dag}\cE(\ket{i}\bra{j})U\ket{j}.
\end{equation}
Since the set of input states and the set of POVM elements each form a complete operator basis,
we can make the following expansions:
\begin{equation}\label{eq: input state expansion}
\ket{i}\bra{j}=\sum_k b^{(ij)}_k\rho_k
\end{equation}
\begin{equation}\label{eq: POVM expansion}
U\ket{j}\bra{i}U^{\dag}=\sum_l c^{(ji)}_l E_l
\end{equation}
Plugging back into Eq.~\ref{eq: entanglement fidelity 2}, we get
\begin{align}\label{eq: entanglement fidelity 3}
 F(\cE,U)&=\frac{1}{d^2}\sum_{ij}\sum_{kl}b^{(ij)}_k c^{(ji)}_l \mathrm{Tr}(E_l\cE(\rho_k))\notag\\
&= \frac{1}{d^2}\sum_{ij}\sum_{kl}b^{(ij)}_k c^{(ji)}_l \mathrm{Pr}(E_l|\rho_k)
\end{align}
and therefore $a_{kl}=\sum_{ij}b^{(ij)}_k c^{(ji)}_l$. It remains to determine the coefficients $b^{(ij)}_k$ and $c^{(ji)}_l$.
The set $\{E_l\}_l$ is an overcomplete basis, so there is a degenerate set of solutions to Eq.~\ref{eq: POVM expansion}.
We proceed as follows:  let $\ket{\rho_k}\rangle$ be a vectorization
of $\rho_k$ and define the superoperator $\cS=\sum_k \ket{\rho_k}\rangle\langle\bra{\rho_k}$.
Then the vectorized \textit{dual basis} density matrix $\tilde{\rho_k}$ corresponding to $\rho_k$
is defined by $\ket{\tilde{\rho_k}}\rangle = \cS^{-1}\ket{\rho_k}\rangle$.
The dual basis POVM elements $\tilde{E_l}$ are defined analogously.
Let $\ket{ij}\rangle$ be a vectorization of $\ket{i}\bra{j}$. Then
\begin{align}
\ket{ij}\rangle &= \cS\cS^{-1}\ket{ij}\rangle\notag\\
 &= \sum_k \ket{\rho_k}\rangle\langle\bra{\rho_k}\cS^{-1}\ket{ij}\rangle\notag\\
&= \sum_k \ket{\rho_k}\rangle\langle\bra{\tilde{\rho_k}}ij\rangle\rangle,
\end{align}
and therefore a solution to Eq.~\ref{eq: input state expansion} is given by
$b^{(ij)}_k = \langle\bra{\tilde{\rho_k}}ij\rangle\rangle = \bra{j}\tilde{\rho_k}\ket{i}$.
Similarly, a solution to Eq.~\ref{eq: POVM expansion} is given by $c^{(ji)}_l = \bra{i}U^{\dag}\tilde{E_l}U\ket{j}$.
Plugging into Eq.~\ref{eq: entanglement fidelity 3} and after some simplification,
we find that the coefficients in the linear estimator are given by
\begin{equation}
a_{kl} = \frac{1}{d^2}\mathrm{Tr}(U\tilde{\rho_k}U^{\dag}\tilde{E_l}).
\end{equation}
The linear fidelity estimator is a consistent and unbiased estimator, meaning that $\hat{F}_L$ converges to $F(E,U)$ in the limit of infinite trials per experiment and the expectation of $\hat{F}_L$ is $F(E,U)$. However, $\hat{F}_L$ has a larger variance than $\hat{F}_{ML}$, in part
due to effects that occur at the boundary of quantum state space. We generate
2000 non-parametric bootstrapped data sets to obtain error bars on the linear fidelity estimate.
The distribution of bootstrapped linear fidelity estimates is shown in Fig.~\ref{fig:boot_fids}.
As $\hat{F}_L$ is unbiased, its value for the experimental data approximately equals the mean of the bootstrapped distribution. Therefore the basic bootstrap confidence interval matches the interval between the corresponding quantiles of the distribution.
We obtain a 95\% basic bootstrap confidence interval of \fidledstwo for $\hat{F}_L$, which contains the ML entanglement fidelity. 
The results from the ML estimation and the linear estimation of the two data sets are summarized in
Table~\ref{tab:analysis_sum}.

\paragraph*{\paraptm}

The ML-estimated process shown in Fig.~3 of the main text is in the Pauli transfer matrix
representation. Let $\{P_i\}_{i=0}^{d^2-1}$ be an operator basis of $n$-qubit Pauli operators, where $d=2^n$, and $P_0=I$.
Then the \textit{Pauli transfer matrix} $\mathcal{T}$ has entries $\mathcal{T}_{ij}=\frac{1}{d}\mathrm{Tr}\big(P_i\cE(P_j)\big)$.
The CPTP constraint on $\cE$ results in the properties that $\mathcal{T}_{00}=1$, $\mathcal{T}_{0j}=0$ for $j>0$, and
$-1\le\mathcal{T}_{ij}\le 1$ for all $i$,$j$. The Pauli transfer matrix elements are related to the Choi matrix by
\begin{equation} 
\mathcal{T}_{ij}= \mathrm{Tr}\big(\chi(P_j^{\intercal}\otimes P_i)\big).
\end{equation}

\paragraph*{\paralrt}

As a further consistency check, we perform a likelihood ratio test to investigate whether a CPTP map acting on a
two-qubit state space is a good model for the observed data. The general problem of deciding between models to
fit data is called model selection. 
For a recent reference on the use of model selection in quantum tomography, see \cite{scholten_behavior_2018}.
A model $\mathcal{M}$ is a parametrized set of probability distributions. Given two nested models $\mathcal{M}_0\subset\mathcal{M}_1$,
the likelihood ratio test allows one to decide whether or not to reject the null hypothesis
that the observed data is sampled from a distribution in the model $\mathcal{M}_0$. In our case, let $\mathcal{M}_0$ denote the model of
all probability distributions that could result from the application of a single CPTP map $\cE$ on each trial during the process tomography protocol.
Let $\mathcal{M}_1$ denote the fully unrestricted model, that is,
the set of all 144 independent probability distributions (one distribution for each combination of state preparation and measurement basis) on 4 elements. The \textit{log-likelihood ratio statistic} $\lambda$
is defined by
\begin{equation}\label{eq: log-likelihood ratio statistic}
    \lambda = 2(\cL(\mathcal{M}_1) - \cL(\mathcal{M}_0)),
\end{equation}  
where $\cL(\mathcal{M}_0) = \mathrm{max}_\mathrm{\cE}(\cL(\cE))$ 
is the maximum of the log-likelihood function defined in Eq.~\ref{eq: loglike},
and $\cL(\mathcal{M}_1)=\sum_{kl}n_{kl}\ln(p_{kl})$ is the maximum log-likelihood of the fully unrestricted model given the observed data.

\begin{figure}[h]
	\centering
	\includegraphics[width = 1.0\textwidth]{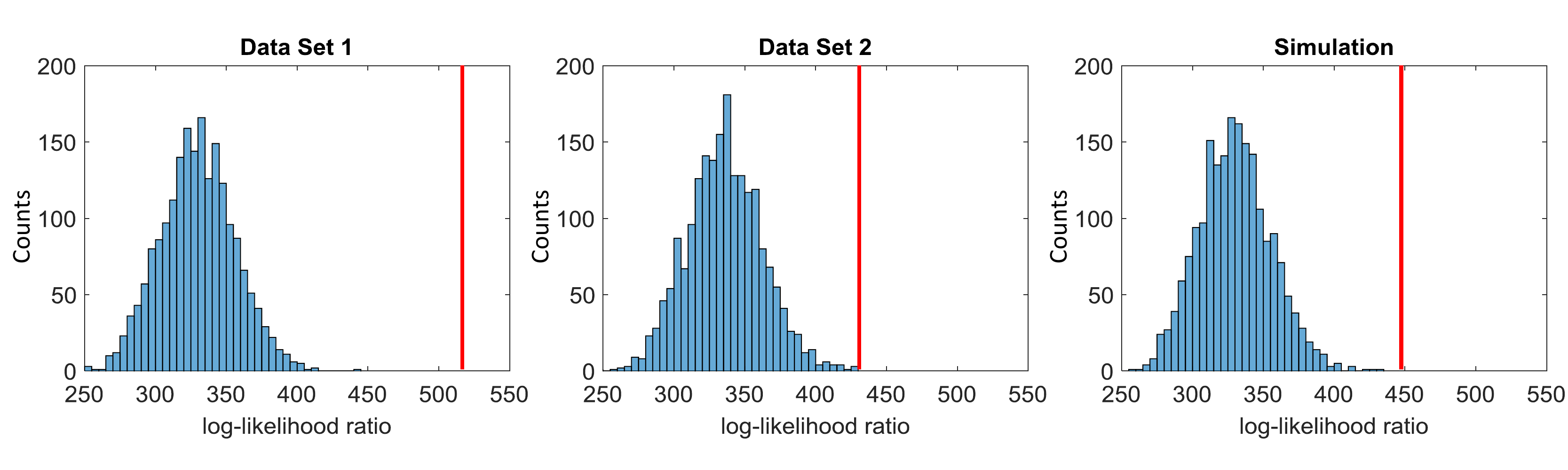}
	\caption{\textbf{Likelihood ratio tests.} Comparison of the log-likelihood ratio
		statistic $\lambda$ defined in Eq.~\ref{eq:
			log-likelihood ratio statistic} with the distribution of values of $\lambda$ obtained from 2000 parametric bootstrap resamples from the ML process. The deviation of the experimental value of $\lambda$ from the distribution are 7.1, 3.6, and 4.6 standard deviations for dataset 1, dataset 2, and a simulated data set, respectively.
		The simulated data set results from a simulation of the experiment that models systematic errors caused by drifts in the \beshort single-qubit rotation angle, as explained in the text.
	}
	\label{fig: log-likelihood ratio}
\end{figure}

Likelihood ratio tests often assume that $\lambda$ has a chi-squared distribution, but because of boundary effects, this assumption is typically not true in quantum tomography experiments \cite{scholten_behavior_2018}.  Rather than using a chi-squared distribution, our likelihood ratio test compares the value of $\lambda$ computed from the experimental data with
the distribution of values for $\lambda$ obtained from the bootstrapped data sets. Roughly, if the experimental value of
$\lambda$ is near the center of the bootstrapped distribution, 
then there is no statistical evidence for rejecting $\mathcal{M}_0$.
We quantify model discrepancy as $(\lambda-\bar{\lambda})/\sigma$, where $\lambda$ is obtained from the original data, and $\bar{\lambda}$ and $\sigma$ are the mean and standard deviation of the log-likelihood ratio statistic from the bootstrapped data sets.
The results of the likelihood ratio test are shown in Fig.~\ref{fig: log-likelihood ratio} . We observe model discrepancy at the level of
7.1 and 3.6 standard deviations for dataset 1 and 2, respectively. This indicates that
the data is inconsistent with the null model $\mathcal{M}_0$:
the application of a single CPTP map to the specified input states followed by measurement by the inferred POVM does not fit the data as well as the fully unrestricted model.
Such a discrepancy could be caused by context dependence, that is, a systematic
dependence of the applied process on external variables \cite{rudinger_probing_2018}.

\begin{figure}[h]
	\centering
	\includegraphics[width=0.8\textwidth]{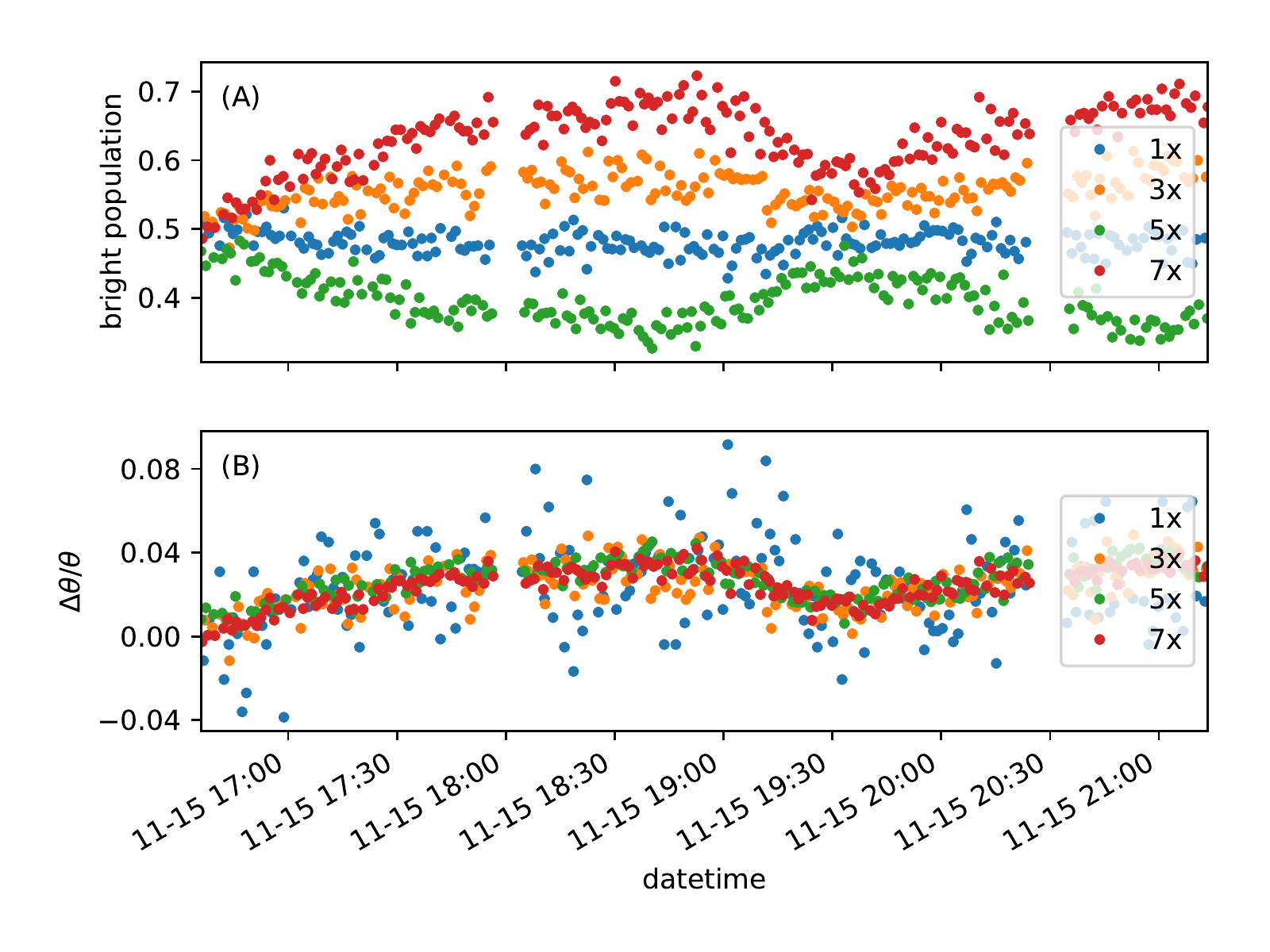}  
	\caption{\textbf{Drifts of single-qubit-rotation pulses.} The rotation-angle drifts of single-qubit-rotations on \beshort ions are measured by
		applying 1 (blue), 3 (orange), 5 (green), and 7 (red) $\pi/2$ pulses.  The odd number of single-qubit-rotation pulses ideally excite the population to around $0.5$, a point where the slope of the population as a function of rotation angle is maximal, providing maximum sensitivity to rotation angle changes. Each single $\pi/2$ pulse takes approximately 10\,\us with about 40\,\us gap between the pulses.  Data taken on November 15, 2018.
		(A) Bright population after single-qubit rotation pulses.  
		(B) Fractional changes of rotation angles derived from measured population. (See text)
	}
	\label{fig:becocarr_drift}
\end{figure}

A known systematic error present in the experiment is the drifts of Rabi rates for single-qubit rotations
on the \beshort ions.
To investigate whether such drifts may be responsible for the observed model discrepancy,
we simulated a data set with correlated over-rotation errors on the state preparation and measurement pulses
as well as the single-qubit rotations when implementing the \cnot gates on \bmpair pairs shown in Fig.~\ref{fig:composite_gates}C.
The magnitudes of the errors in the simulation drift according to the pattern observed in
Fig.~\ref{fig:becocarr_drift}.
A likelihood ratio test on the simulated data set yields a discrepancy at the level of
4.6 standard deviations, as shown in Fig.~\ref{fig: log-likelihood ratio}.
We conclude that a large portion of the model discrepancy observed in the experimental data can be explained by drifts in the rotation angles. 


To verify the existence  of rotation-angle drifts for single-qubit rotations on \beshort ions in
the experimental setup, we performed a separate investigation in which we monitored the population of \beshort ions after applying various numbers of 
$\pi/2$ pulses as shown in Fig.~\ref{fig:becocarr_drift}A. The deviation from the starting population was converted to fractional changes in
rotation angles $\delta\theta/\theta$ (Fig.~\ref{fig:becocarr_drift}B), where $\theta$ is the target pulse area, and $\delta\theta$ is the deviation of the actual pulse area from the target value. A maximum fractional change of up to $4\%$ was observed over a period of 4 hours in the measurement shown. 
The drifts are mainly caused by the relative power fluctuation between the two spatially overlapped laser beams used to drive the
co-propagating carrier, while the total power of the two is actively stabilized. 
The power ratio between the two beams was about $1:3$ at the time of the measurement. Balancing the power
ratio between the two beams reduced the drifts of Rabi rates,  but the
drifts at the percentage level remained. This can be improved  by   actively stabilizing the beat-note amplitude  of
the two beams in  future experiments.
 
The likelihood ratio test helped 
identify a time-varying experimental imperfection that caused the experiment to depart from our model of a static process matrix.  
We believe this illustrates the importance of performing model consistency checks in addition to tomography when characterizing quantum processes.
 
\paragraph*{\paraes}  
We use a depolarizing model to estimate the total error of the QGT process (referred to as the algorithm in the following). Assuming
a depolarizing error $\epsilon_i$ for a constituent process $\hat{U}_i$ of the algorithm, the density matrix after the process becomes
\begin{equation}
	\rho \rightarrow (1-\epsilon_i) \hat{U}_i\rho \hat{U}_i^\dagger  + \epsilon_i \cdot \hat{I} / d_i
\end{equation}
where $d_i$ is the dimension of the Hilbert space for the process. Stepping through the full algorithm and applying a depolarizing error with the same magnitude  as experimentally determined for each constituent process (which may or may not be depolarizing in reality) allows us to compute an approximate
 density matrix after the complete QGT process, and therefore derive the process matrix of this model process and its fidelity with respect to an ideal
\cnot.

The error sources  we consider here are SPAM errors
on each qubit, the gate errors of each composite gate (Fig.~\ref{fig:composite_gates}), the error due to decoherence of the \mgshort ions, and  depumping errors induced by stray resonant light. 

\begin{figure}[h]
	\centering
	\includegraphics[width=1.0\textwidth]{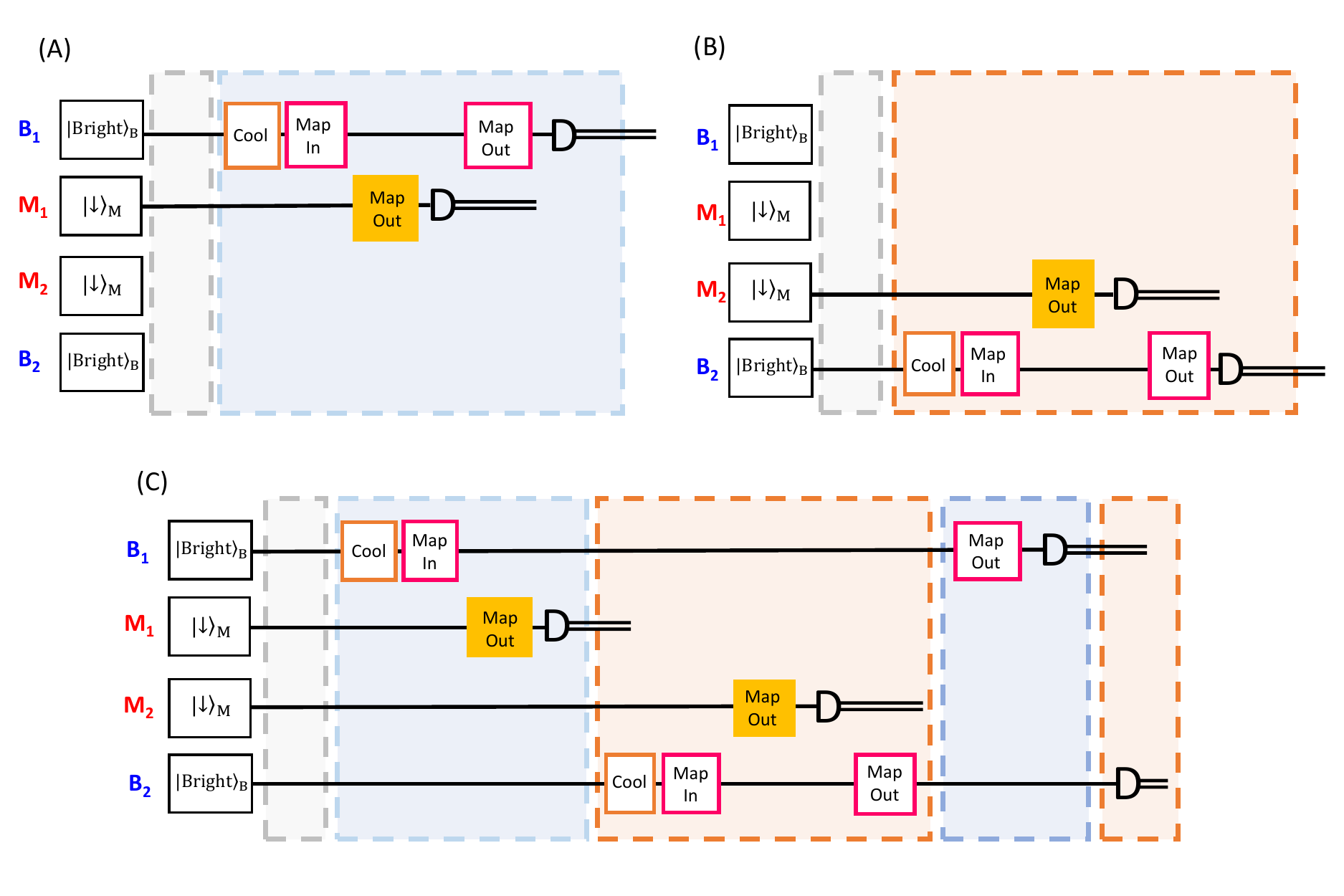}  
	\caption{\textbf{Experiments for determining SPAM and depumping errors.} (A) and (B) Sequence for determining the SPAM errors for \beshort and \mgshort ions
		detected in the double well potential.
		These are also the reference
		experiments for determining the stray-light-induced
		depumping errors. (C) To determine the stray-light-induced depumping error, cooling, state preparation, and measurement pulses are applied in the same order as in the QGT
		sequence to conduct SPAM measurements for all four qubits. These measurements contain the SPAM errors plus
		the stray-light-induced depumping errors. The difference between this and the reference experiments in (A) and (B) gives
		the stray-light-induced depumping error. 
	}
	\label{fig:spam_sequence}  
\end{figure}

The SPAM errors for the individual qubits are taken from  SPAM diagnosis experiments (see Fig.~\ref{fig:spam_sequence}A and \ref{fig:spam_sequence}B) interleaved
with the tomography measurements.  
In the SPAM diagnosis experiments, we  find the probability $\epsilon_{X, \uparrow}$ of measuring an ion in $\ket{\downarrow}_{X}$ after preparing it in the $\ket{\uparrow}_{X}$ state, where $X=$\,\beone, \betwo, \mgone, or \mgtwo. We also find the probability $\epsilon_{X, \downarrow}$ of measuring an ion in $\ket{\uparrow}_{X}$ after preparing it in the $\ket{\downarrow}_{X}$ state.
The ``Map In'' and ``Map Out'' pulses for these measurements are described in section \paraspam. We use the mean value  $\bar{\epsilon}_{X} = \frac{\epsilon_{X,\uparrow} + \epsilon_{X,\downarrow}}{2}$ to estimate the SPAM errors for each  qubit in the QGT sequence.
In addition, we also measured the SPAM errors of two \mgshort ions in a static well, using only microwave pulses for ``Map In'' and ``Map Out''. These results are listed in Tab.~\ref{tab:spam} and used later to account for errors in individual gates.

We use the Bell-state
infidelities from each composite two-qubit gate with the contribution from SPAM errors subtracted as a representative
estimate of  their process infidelities. When estimating the Bell-state infidelity by measuring qubit populations and the contrast of  parity oscillations \cite{sackett_experimental_2000}, each qubit contributes an amount $\epsilon_{X,\mathrm{Bell}} = \frac{3}{2}\bar{\epsilon}_{X}$ to the observed Bell-state infidelity \cite{ballance_high-fidelity_2014}. After subtracting the contribution from SPAM errors, we estimate an infidelity  of $0.040(9)$ for the Bell-state-generating gate $\hat{F}$,
and $0.030(9)$ and $0.03(1)$ for the mixed-species \cnot gates. 

The error from \mgshort ion decoherence is estimated from the coherence time of a Ramsey
experiment on a single \mgshort ion. Here we account for the fact that the two \mgshort ions are in the Bell state for a
duration of $4.2\,\mathrm{ms}$ and that a coherent superposition needs to be preserved in \mgtwo for a
further $3.6\,\mathrm{ms}$. We model this decoherence error as equivalent to the contrast
reduction in the  Ramsey sequence on a single \mgshort ion with a wait time of $2\times4.2 +
3.6=12\,\mathrm{ms}$. From the 	 $1/e$ coherence time of
$140(30)\,\mathrm{ms}$, 
we estimate the error from
 decoherence of the two \mgshort ions in the algorithm to be  $0.007(3)$.

We further consider the errors from stray-resonant-light-induced
depumping on ions outside of the LIZ. This is quantified as the difference between
two SPAM experiments. A  reference experiment measures the SPAM errors for one pair of \bmpair ions by
shuttling the pair to the LIZ and performing the measurement there
(Fig.~\ref{fig:spam_sequence}A,B). In a separate experiment, we perform shuttling, cooling, state preparation, and detection in the same order
as in the QGT experiment to measure the SPAM error
plus the stray-light-induced depumping error  (Fig.~\ref{fig:spam_sequence}C).
The difference of the two experiments is taken as the depumping error. We  obtain errors of $0.011(4)$ for \mgone (due to the detection of \mgtwo) and $0.012(3)$ for \betwo (due to
cooling and repumping of \beone).

Based on the following assumptions and estimates, we can reject the hypothesis 
that the stray-resonant-light-induced depumping errors are induced
solely by light scattering from the ions in the LIZ. 
Assuming a scattering rate of $R_1=\gamma_\mgsubs/2$ for the \mgone ion at the LIZ from a detection pulse with a duration of $\tau=200\,\us$, the \mgtwo ion away from the LIZ experiences an intensity of $I_2 \simeq \frac{\gamma_\mgsubs}{2}\cdot\frac{hc}{\lambda_\mgsubs\cdot 4\pi l^2}$ from the scattering, where $\gamma_\mgsubs$ is the decay rate of the upper state of \mgshort ions, $h$ is the Planck's constant, $c$ is the speed of light, $\lambda_\mgsubs$ is the wavelength of \mgshort resonant light, and $l=340\,\um$ is the smallest distance between two minima of the double well. The \mgtwo ion will scatter at a rate $R_2=\frac{I_2}{hc/\lambda_\mgsubs}\sigma_\mgsubs$ where $\sigma_\mgsubs\simeq\frac{3}{2\pi}\lambda_\mgsubs^2$ is the absorption cross section of \mgshort ions at resonance. This implies a scattering probability of about $6\times10^{-4}$ for a detection period of $200\,\us$, which is much lower than the observed error  rate of approximately one percent.

Depumping errors on the order of one percent indicate that on average at least 0.01 photons are scattered in one experiment due to fluorescence detection at the LIZ. In comparison,  $\sim\,$$1000$ resonant
photons are scattered by an ion in the bright state in the LIZ. This is approximately equivalent to an 
intensity ratio of $\sim\,$$0.01/1000 = 10^{-5}$ between the resonant light intensity outside of the
LIZ compared to the light intensity in the LIZ. Reaching the
goal of $10^{-4}$ error rates for practical fault tolerant quantum error correction would then require an intensity
ratio of $\sim\,$$10^{-4}/1000=10^{-7}$. This presents a significant engineering requirement in future ion traps,
but one that could be removed by completely eliminating the use of qubit resonant light and instead
restricting all resonant operations to  separate species ancillas \cite{tan_multi-element_2015}. After any previous resonant interaction of a nearby ancilla ion has taken place, affected ancillas can be prepared again before use in the algorithm.
 
Inserting all errors from Table~1 in the main text as depolarizing errors into the	 model, we derive a total infidelity of $0.12(1)$ for the QGT algorithm. We derive the uncertainty of the total infidelity
by applying a bootstrap to the depolarizing model. 
The fidelity determined from the depolarizing model is near the upper limits of the 95\% confidence intervals obtained by both ML and linear estimation, indicating that the major error sources are reflected in this simplified error propagation model.

Other error sources, such as drifting experimental parameters, are not included in the depolarizing model. 
In addition, the drifts ins preparation and measurement pulses can  cause bias in the estimate of the process fidelity. For example, we determine by 
 simulation that rotation-angle drifts  of single-qubit rotations on \beshort ions, uniformly distributed up to $5\%$,
lead to an under-estimation of process fidelity by $\sim\,$$1\%$.

\begin{table}[h]
	\caption{{\bf Detailed steps of the QGT algorithm with the approximate durations of each step.} 
		The majority of the time is spent on cooling and shuttling the ions. DC: Doppler
		cooling, SP:
		state preparation, SBC:
		sideband cooling. The additional \mgshort ion detections in steps E and F are for diagnostics and are not part of the QGT algorithm.}
	\label{tab:operations}
	
	\begin{center}
		\begin{tabular}{clcc}
			\toprule
			Step & Description                & Operation & Duration \\
			\midrule \midrule
			\multirow{5}{*}{A} & Optical pumping and crystal initialization & \multirow{2}{*}{-} & \multirow{2}{*}{$3.2\,\mathrm{ms}$} \\  
			& \quad of the \fouriontxt & & \\
			& DC on \beshort and \mgshort and SP of \mgshort ions         
			& -             & $1.3\,\mathrm{ms}$ \\
			& SBC on \beshort ions						  &       -             &  	$5.3\,\mathrm{ms}$ \\
			& \mgone--\mgtwo Bell-state generation     			  &       $\hat{F}$   	& 	$220\,\us$\\
			\cline{1-4}
			\multirow{2}{*}{B}    & Separation of \fouriontxt into \beone--\mgone &      \multirow{2}{*}{-}   & 
			\multirow{2}{*}{$570\,\us$}
			\\
			&  \quad and \mgtwo--\betwo in double well & & \\ 
			\cline{1-4}
			\multirow{4}{*}{C} & Shifting double well    &  - 	&  $230\,\us$\\    
			& DC, SBC, and SP of \beone    &  - 	&  $2.2\,\mathrm{ms}$\\
			& \cnot on \beone--\mgone                                     &         \cnot &
			$280\,\us$
			\\
			& Map out and detection of \mgone                       &         -  &  $650\,\us$\\
			\cline{1-4}
			\multirow{7}{*}{D} & Shifting double well & - & $460\,\us$\\  
			& Cooling and SP of \betwo  &          - &  $2.2\,\mathrm{ms}$
			\\
			& Conditional rotation on \mgtwo                       &         $\hat{R}(\pi, 0)$  & $25\,\us$\\
			& \cnot on \betwo--\mgtwo                      &         \cnot &  $280\,\us$ \\
			& Rotation on \mgtwo                       &         $\hat{R}(\pi/2, -\pi/2)$  & $15\,\us$ \\
			& Map out and detection of \mgtwo                       &         -  &  $650\,\us$ \\
			& Map out \betwo                       &         -  &  $220\,\us$ \\
			\cline{1-4}  
			\multirow{5}{*}{E}     & Shifting double well & - & $460\,\us$ \\
			& DC on \mgone & - & $200\,\us$\\   
			
			& Conditional phase shift on \beone  & $\hat{R}_\mathrm{Z}(\pi)$ &  $100\,\mathrm{ns}$ \\
			& Map out and detection of \beone            &       -    & $540\,\us$ \\
			& Detection of \mgone   & & $180\,\us$ 	 			        \\
			\cline{1-4}
			\multirow{5}{*}{F} & Shifting double well & - & $460\,\us$\\
			& DC on \mgtwo & - & $200\,\us$\\   
			& Detection of \betwo &      -     &  $300\,\us$  \\
			& Detection of \mgtwo & - & $180\,\us$ \\
			& Recombination              &       -    & $800\,\us$\\  
			\bottomrule
		\end{tabular}
	\end{center}
\end{table}

\begin{table}[h]
	\caption{{\bf Entanglement fidelities and confidence intervals.} Entanglement fidelities and confidence intervals of the teleported \cnot gate are determined using maximum likelihood (ML) estimation
		and a linear estimator.
	} 
	\label{tab:analysis_sum}
	
	\begin{center}
		\begin{tabular}{lcc}
			& dataset 1 & dataset 2 \\
			\midrule
			Fidelity of ML process	& $0.858$   & $0.851$ \\
			Fidelity ($68\%$ CI from ML) &  $[0.858, 0.871]$ & $[0.852, 0.866]$ \\
			Fidelity ($95\%$ CI from ML)
			& $[0.852, 0.878]$ & $[0.845, 0.872] $ \\
			Fidelity from linear estimator $\hat{F}_L$ & $0.871$ & $0.866$ \\
			Fidelity ($68\%$ CI from linear estimator) & $[0.860, 0.882]$ & $[0.856, 0.878]$ \\
			Fidelity ($95\%$ CI from linear estimator) & $[0.851, 0.892]$ & \fidledstwo \\
		\end{tabular}
	\end{center}
\end{table}

\begin{table}[h]
	\caption{{\bf SPAM errors measurements.}
		The uncertainties are determined from the standard deviation of multiple measurements. 
	}
	\label{tab:spam}
	
	\begin{center}
		\begin{tabular}{cccccc}
			& $\epsilon_{X,\uparrow}$ $(10^{-2})$ & $\epsilon_{X,\downarrow}$ $(10^{-2})$  &  $\bar{\epsilon}_{X}$ $(10^{-2})$ & $\epsilon_{X,\mathrm{Bell}}$ $(10^{-2})$  \\
			\midrule
			\beone	& 0.5(4)   & 0.4(2)  &  0.5(2) & 0.7(3) \\
			\betwo &  1(1) & 0.4(2) & 0.7(6) & 1.0(9)\\
			\mgone			& 0.2(1) & 1.4(4) & 0.8(2) & 1.2(3) \\
			\mgtwo & 0.2(1) & 1.3(3) & 0.7(2) &   1.1(3) \\
			Both \mgshort ions (microwave) & 0.2(1) & 2.2(4) & 1.2(2) & 1.8(3) \\
		\end{tabular}
	\end{center}
\end{table}


\begin{thebibliography}{10}

	\bibitem{preskill_quantum_2018}
	J.~Preskill, {\it Quantum\/} {\bf 2}, 79 (2018).

	\bibitem{gottesman_demonstrating_1999}
	D.~Gottesman, I.~L. Chuang, {\it Nature\/} {\bf 402}, 390 (1999).

	\bibitem{eisert_optimal_2000}
	J.~Eisert, K.~Jacobs, P.~Papadopoulos, M.~B. Plenio, {\it Physical Review A\/}
	{\bf 62}, 052317 (2000).

	\bibitem{wineland_experimental_1998}
	D.~J. Wineland, {\it et~al.\/}, {\it Journal of Research of the National
		Institute of Standards and Technology\/} {\bf 103}, 259 (1998).

	\bibitem{kielpinski_architecture_2002}
	D.~Kielpinski, C.~Monroe, D.~J. Wineland, {\it Nature\/} {\bf 417}, 709 (2002).

	\bibitem{monroe_large-scale_2014}
	C.~Monroe, {\it et~al.\/}, {\it Physical Review A\/} {\bf 89}, 022317 (2014).

	\bibitem{lin_dissipative_2013}
	Y.~Lin, {\it et~al.\/}, {\it Nature\/} {\bf 504}, 415 (2013).

	\bibitem{moehring_entanglement_2007}
	D.~L. Moehring, {\it et~al.\/}, {\it Nature\/} {\bf 449}, 68 (2007).

	\bibitem{pirandola_advances_2015}
	S.~Pirandola, J.~Eisert, C.~Weedbrook, A.~Furusawa, S.~L. Braunstein, {\it
		Nature Photonics\/} {\bf 9}, 641 (2015).

	\bibitem{bennett_teleporting_1993}
	C.~H. Bennett, {\it et~al.\/}, {\it Physical Review Letters\/} {\bf 70}, 1895
	(1993).

	\bibitem{huang_experimental_2004}
	Y.-F. Huang, X.-F. Ren, Y.-S. Zhang, L.-M. Duan, G.-C. Guo, {\it Physical
		Review Letters\/} {\bf 93}, 240501 (2004).

	\bibitem{gao_teleportation-based_2010}
	W.-B. Gao, {\it et~al.\/}, {\it Proceedings of the National Academy of
		Sciences\/} {\bf 107}, 20869 (2010).

	\bibitem{chou_deterministic_2018}
	K.~S. Chou, {\it et~al.\/}, {\it Nature\/} {\bf 561}, 368 (2018).

	\bibitem{tan_multi-element_2015}
	T.~R. Tan, {\it et~al.\/}, {\it Nature\/} {\bf 528}, 380 (2015).

	\bibitem{noauthor_materials_nodate}
	Materials and methods are available as supplementary materials at the {Science}
	website.

	\bibitem{blakestad_high-fidelity_2009}
	R.~B. Blakestad, {\it et~al.\/}, {\it Phys. Rev. Lett.\/} {\bf 102}, 153002
	(2009).

	\bibitem{blakestad_near-ground-state_2011}
	R.~B. Blakestad, {\it et~al.\/}, {\it Physical Review A\/} {\bf 84}, 032314
	(2011).

	\bibitem{sorensen_quantum_1999}
	A.~S\o{}rensen, K.~M\o{}lmer, {\it Phys. Rev. Lett.\/} {\bf 82}, 1971 (1999).

	\bibitem{lee_phase_2005}
	P.~J. Lee, {\it et~al.\/}, {\it Journal of Optics B: Quantum and Semiclassical
		Optics\/} {\bf 7}, S371 (2005).

	\bibitem{chuang_prescription_1997}
	I.~L. Chuang, M.~A. Nielsen, {\it Journal of Modern Optics\/} {\bf 44}, 2455
	(1997).

	\bibitem{casella_statistical_nodate}
	G.~Casella, R.~L. Berger, {\it Statistical {Inference}\/} (Cengage Learning,
	2001), p. 374.

	\bibitem{blume-kohout_demonstration_2017}
	R.~Blume-Kohout, {\it et~al.\/}, {\it Nature Communications\/} {\bf 8}, 14485
	(2017).

	\bibitem{efron_introduction_1994}
	B.~Efron, R.~J. Tibshirani, {\it An {Introduction} to the {Bootstrap}\/} (CRC
	Press, 1994).

	\bibitem{schmidt_spectroscopy_2005}
	P.~O. Schmidt, {\it et~al.\/}, {\it Science\/} {\bf 309}, 749 (2005).

	\bibitem{slichter_uv-sensitive_2017}
	D.~H. Slichter, {\it et~al.\/}, {\it Optics Express\/} {\bf 25}, 8705 (2017).

	\bibitem{bowler_coherent_2012}
	R.~Bowler, {\it et~al.\/}, {\it Phys. Rev. Lett.\/} {\bf 109}, 080502 (2012).

	\bibitem{walther_controlling_2012}
	A.~Walther, {\it et~al.\/}, {\it Phys. Rev. Lett.\/} {\bf 109}, 080501 (2012).

	\bibitem{roos_experimental_2000}
	C.~F. Roos, {\it et~al.\/}, {\it Physical Review Letters\/} {\bf 85}, 5547
	(2000).

	\bibitem{wan_data_2019}
	Y. Wan, Data for "Quantum gate teleportation between separated qubits in a trapped-ion processor," Version 1, National Institute of Standards and Technology (2019);
	http://doi.org/10.18434/M32056.

	\bibitem{gaebler_high-fidelity_2016}
	J.~P. Gaebler, {\it et~al.\/}, {\it Phys. Rev. Lett.\/} {\bf 117}, 060505
	(2016).

	\bibitem{monroe_resolved-sideband_1995}
	C.~Monroe, {\it et~al.\/}, {\it Physical Review Letters\/} {\bf 75}, 4011
	(1995).

	\bibitem{wan_efficient_2015}
	Y.~Wan, F.~Gebert, F.~Wolf, P.~O. Schmidt, {\it Physical Review A\/} {\bf 91},
	043425 (2015).

	\bibitem{keith_joint_2018}
	A.~C. Keith, C.~H. Baldwin, S.~Glancy, E.~Knill, {\it Phys. Rev. A\/} {\bf 98},
	042318 (2018).

	\bibitem{choi_completely_1975}
	M.-D. Choi, {\it Linear Algebra and its Applications\/} {\bf 10}, 285 (1975).

	\bibitem{Fiurasek_maximum-likelihood_2001}
	J.~Fiurášek, {\it Physical Review A\/} {\bf 64}, 024102 (2001).

	\bibitem{schumacher_sending_1996}
	B.~Schumacher, {\it Physical Review A\/} {\bf 54}, 2614 (1996).

	\bibitem{scholten_behavior_2018}
	T.~L. Scholten, R.~Blume-Kohout, {\it New Journal of Physics\/} {\bf 20},
	023050 (2018).

	\bibitem{rudinger_probing_2018}
	K.~Rudinger, {\it et~al.\/}, {\it arXiv:1810.05651\/}  (2018).

	\bibitem{sackett_experimental_2000}
	C.~A. Sackett, {\it et~al.\/}, {\it Nature\/} {\bf 404}, 256 (2000).

	\bibitem{ballance_high-fidelity_2014}
	C.~J. Ballance, {\it High-{Fidelity} {Quantum} {Logic} in {Ca}$^{+}$\/},
	Thesis, University of Oxford (2014).

\end{thebibliography}
\end{document}